\documentclass{article}

\usepackage{arxiv}

\usepackage[utf8]{inputenc} 
\usepackage[T1]{fontenc}    
\usepackage{hyperref}       
\usepackage{url}            
\usepackage{booktabs}       
\usepackage{amsfonts}       
\usepackage{nicefrac}       
\usepackage{microtype}      
\usepackage{lipsum}
\usepackage{amssymb} 
\usepackage{mdframed}
\usepackage{booktabs}
\usepackage{todonotes}
\usepackage{xcolor}
\usepackage{placeins}

\usepackage{amsmath}
\usepackage{dirtytalk}
\newcommand{\vp}{\color{black}}
\newcommand{\vm}{\color{black}}

\newcommand{\lp}{\color{black}}

\usepackage{soul}
\newcommand{\upd}{\color{black}}
\usepackage{graphicx}
\graphicspath{ {./images/} }
\usepackage{natbib}
\title{Joint Effects of Recommender Systems and Network Structure on the Visibility of Content and Creators}

\author{
 Virginia Morini \\
  Department of Computer Science, University of Pisa \\
  ISTI-CNR, National Research Council of Italy \\
  Pisa, Italy \\
  \texttt{virginia.morini@di.unipi.it} \\
  \And
 Valentina Pansanella \\
  ISTI-CNR, National Research Council of Italy \\
  Pisa, Italy \\
  \texttt{valentina.pansanella@isti.cnr.it} \\
  \And
 Luca Pappalardo \\
  ISTI-CNR, National Research Council of Italy \\
  Scuola Normale Superiore \\
  Pisa, Italy \\
  \texttt{luca.pappalardo@isti.cnr.it} \\
  \And
 Dino Pedreschi \\
  Department of Computer Science, University of Pisa \\
  Pisa, Italy \\
  \texttt{dino.pedreschi@unipi.it} \\
  \And
 Giulio Rossetti \\
  ISTI-CNR, National Research Council of Italy \\
  Pisa, Italy \\
  \texttt{giulio.rossetti@isti.cnr.it} \\
}
\makeatletter
\renewcommand{\@date}{}
\makeatother

\begin{document}

\maketitle

\begin{abstract}
Social media feeds allocate visibility by ranking content within users' social
networks. Yet how ranking logic and network structure jointly shape who and what
becomes visible remains largely understudied. We
address this through agent-based simulations in YSocial, a social media virtual
twin, comparing \textcolor{black}{six} recommendation strategies
{\upd on a scale-free network and {\color{black} varying the topology and the
main simulation parameters to assess robustness}}.
{\upd Recommendation strategy primarily determines the pattern of visibility
allocation, while network structure shapes how exposure is distributed across
creators.} Popularity ranking concentrates exposure on a small subset of content
and creators and couples early engagement to later exposure, whereas
collaborative filtering, {\upd in our controlled setting,} keeps nearly all
active content and every creator in circulation. Crucially, the same popularity
signal produces different inequalities depending on the candidate pool it ranks:
applied globally, it selects a few contents largely irrespective of their
creators' network position; applied within the follower graph, it
{\upd increases content coverage} but redirects visibility toward already highly
connected creators. {\upd This amplification is stronger in degree-heterogeneous
networks, while additional clustering changes little.}
{\upd These patterns persist across settings, but their magnitude depends on
feed size, content lifetime, simulation duration, and system scale; at larger
scales the follower graph no longer moderates popularity-driven concentration
and reinforcement.} Visibility should therefore be evaluated across content,
creators, network position, and temporal reinforcement, {\upd and across the
conditions under which recommendation operates. Because these are entangled in
deployed platforms, observed inequalities should not be read as effects of the
recommender alone.}
\end{abstract}

\noindent\textbf{Code} --- 
\url{https://anonymous.4open.science/r/icwsm-recsys-visibility-7864/}

\noindent\textbf{Data} --- 
\url{https://tinyurl.com/YSocialRecsysSimulationData}

\section{Introduction}
\label{sec:intro}
Recommender systems have a central role in the attention economy, allocating visibility across competing contents and creators \cite{heitmayer2025second,baeza2021attention}.
This is evident on social media platforms, where algorithmically curated feeds have become the main access point to content and an instrument for user engagement and retention \cite{lewandowsky2022technology}.

In this ecosystem, recommender systems determine which content and creators are visible, how often they are shown, and how exposure is distributed across the platform \cite{ionescu2025visibility}. 
This exposure allocation is not neutral: {by ranking some content above others, recommender systems amplify the visibility of users, topics or opinions, while reducing the visibility of others \cite{huszar2022algorithmic}.}

Prior work has linked {the use of recommendation systems {\lp on social media platforms} to the emergence of unintended} outcomes such as selective exposure, filter bubbles, polarization, and radicalization \cite{bakshy2015exposure,bartley2021auditing,haroon2023auditing,hosseinmardi2024causally}. 
Empirical studies have found that recommendation algorithms concentrate attention on prominent accounts or high-popularity content \cite{huszar2022algorithmic,ye2025auditing}, but they also increase aggregate diversity in some settings \cite{zhou2010the} or deamplify niche content when user utility is considered \cite{ribeiro2023amplification}. 
Therefore, the central question is not whether recommender systems amplify visibility, but \textit{how} different ranking {strategies} allocate exposure across different entities.

Social media platforms have a further layer of complexity because their recommender systems operate on top of a social network. 
Even without algorithmic ranking, {social networks} give highly connected users larger audiences and more opportunities to accumulate attention \cite{zhu2016attention}. 
Friendship and following relationships also constrain the content available for ranking: on Facebook, the composition of users' friend networks is a major filter of political exposure, with algorithmic ranking adding a further layer of selection \cite{bakshy2015exposure}. 
Network-mediated affordances, such as reposts and reshares, provide another visibility channel, expanding the reach of content beyond its original audience and {shaping} what users see in their feeds \cite{guess2023reshares}. 

{Therefore, visibility on social media platforms is shaped by the interaction between ranking algorithms and social network structure: recommender systems amplify advantages present in the social network, and the social network determines which content and creators are best positioned to benefit from ranking signals.
Much of the existing literature examines visibility through the lens of fairness or exposure-mitigation objectives \cite{ge2010beyond, stray2026prosocial}, while the systemic impact on visibility arising from the \textit{interaction} between recommender systems and social networks remains underexplored.} 

{\lp In this work, we address this gap by studying \textit{how the interplay 
between recommendation {systems} and social network structure shapes visibility in social media feeds}.
{We analyze concentration and coverage, {\upd structural advantages}, and popularity reinforcement to assess whether:} 
\textit{i)} different recommendation strategies redistribute or concentrate 
visibility across content and creators; 
\textit{ii)} network {\upd structure} modulates the effects of ranking on creator visibility; 
\textit{iii)} recommendation systems couple early engagement with subsequent exposure.
}

{\lp
To address this goal, we use YSocial, a social media virtual twin that jointly models recommender systems and social network structure~\cite{rossetti2026ysocial}. 
It enables experiments in which a population of agents creates, consumes, and reacts to content while different recommender systems determine which content is shown. 
YSocial provides access to information that is typically \textit{unobservable} on user feeds, including content that was eligible for recommendation but never exposed.
{\upd This enables} a system-level analysis of visibility and allows us to attribute differences in visibility to specific {\upd characteristics of recommender systems and social network structure} rather than to confounding factors.
Accordingly, our analysis is a stylized study of these mechanisms, not an attempt to reproduce the behavior of any deployed platform.
}


We compare six {\lp prototypical recommender systems} deliberately simplified to isolate common ranking strategies rather than reproduce opaque real-world pipelines. 
We organize our analysis around \textit{global recommender systems}, which rank content from all users in the platform, and \textit{network-aware recommender systems}, which prioritize content from followed accounts.
Each system is evaluated against a reverse-chronological baseline, a minimal and widely adopted ordering {strategy} in social media \cite{huszar2022algorithmic, guess2023social}, {\vp on a Barabási--Albert (BA) network \cite{barabasi1999emergence}.}
{\upd We then assess the robustness of these results along two dimensions: we replace the underlying social network with topologies differing in degree heterogeneity
and clustering, and we vary one experimental factor at a time, covering feed size,
content lifetime, observation horizon, and system scale.}
{\vp We then simulate a population of users interacting with the platform over a fixed time horizon.} Finally, we analyze how each recommender system shapes the distribution of visibility across content and creators.

Our results {\lp reveal four main patterns}:
\begin{itemize}
 \item {\lp \textit{Recommendation logic determines who and what gains visibility}: 
 Popularity ranking concentrates exposure on a small subset of content and creators; collaborative filtering keeps nearly all active content and all creators in circulation; and multi-signal ranking produces intermediate outcomes.

\item \lp \textit{Network-aware recommendation links creator visibility to network position}: 
Prioritizing content from followed accounts broadens content circulation relative to global ranking. 
When combined with popularity, however, it shifts visibility toward high-degree creators -- an advantage not produced by global popularity alone.
} 

\item {\lp \textit{Popularity ranking turns early engagement into lasting visibility}: 
Even with random reactions, early engagement predicts subsequent exposure almost perfectly under global popularity ranking, more weakly when popularity is restricted to followed accounts, and not at all under collaborative filtering.
}

{\lp \item \textit{These qualitative patterns persist across simulation settings}: 
Their magnitude varies with feed size, content lifetime, simulation duration, and system scale, but the main distinctions among recommendation mechanisms remain consistent across the tested configurations and network topologies.
}

\end{itemize}

The remainder of the paper is organized as follows. 
We first review related work, then describe the simulation framework, 
recommendation strategies, and visibility metrics. 
We then present results and conclude with a discussion of findings and limitations.

\section{Related Works}
\label{sec:rw}
Recommender systems are studied both as models to optimize, typically for accuracy, engagement, or other target metrics, and as socio-technical systems whose effects extend beyond prediction quality. 
A large body of work has introduced beyond-accuracy metrics, including diversity, novelty, and fairness \cite{ge2010beyond}, while multidisciplinary research in computer science, human--AI interaction, sociology, and complex systems increasingly evaluates their consequences for the social systems they participate in \cite{pedreschi2025human, pappalardo2024survey}. 
We position our work within this latter perspective.

\noindent\textbf{Methods.} The impact of recommender systems in social media has been studied through different methodologies. 
The gold-standard approach is represented by controlled studies, such as A/B tests or randomized trials \cite{huszar2022algorithmic, bakshy2015exposure}. 
{However, such studies remain rare due to data access restrictions and the difficulty of conducting them.} 
External audits and observational studies offer alternatives \cite{ribeiro2020auditing, bouchaud2023algorithmic}, but remain limited by the lack of access to candidate sets, ranking {strategy}, and \say{invisible data}.
Simulation approaches offer a complementary strategy: they allow researchers to control system components, isolate design effects, and study long-term dynamics that are difficult to observe empirically \cite{epstein2012generative, adomavicius2021understanding}. 
More recent frameworks move toward {virtual twins}, reproducing platform primitives such as creating content (i.e., posts and comments), following, reacting, and feed construction, sometimes with LLM-enhanced agents \cite{rossetti2026ysocial, lucherini2021t}.

\noindent\textbf{Recommender Systems and Visibility in Social Media.} 
Prior work shows that different ranking {strategies} allocate visibility in different ways, and that their effects may be mediated by an underlying social {network}. 
The main baseline used to assess the impact of recommender systems is reverse-chronological ranking \cite{huszar2022algorithmic, bartley2021auditing, guess2023reshares, guess2023social, nyhan2023like, perra2019modeling, bouchaud2023algorithmic, bouchaud2024skewed}: although arguably neutral \cite{metzler2024social}, it is considered a minimal form of algorithmic ranking necessary to retrieve relevant information. 
Random recommendation is also commonly employed as a baseline \cite{perra2019modeling, gausen2022using, fabbri2022exposure, ferrara2022link}.
By contrast, recommendations driven directly by popularity have rarely been studied in the social media ecosystem. 
Yet recent survey work shows that recommender systems frequently over-expose popular items, creating popularity bias and reinforcement effects over time \cite{klimashevskaia2024survey}; popularity signals are therefore an important mechanism to isolate and study.
A few works directly model popularity-based recommendations
\cite{bartley2024impact,canamares2014exploring,gausen2022using,lucherini2021t,rossi2021closed},
showing that they can amplify engagement or information diffusion and reinforce exposure distortions.
Germano et al.~\cite{germano2019few} model a dynamic popularity-based ranking in which clicks increase future rank and exposure, showing that popularity feedback can generate concentrated visibility.
The effects of collaborative filtering in social media have mainly been studied in the context of link recommendations impact on minority visibility \cite{ferrara2022link, fabbri2020effect, fabbri2022exposure}.
Despite not directly rewarding popularity, these algorithms are known to have popularity bias: over-exposing popular items while limiting long-tail visibility \cite{abdollahpouri2020multi}; this is however tightly linked to user preferences and behavior. 

\noindent\textbf{Recommender Systems and Network Structure.}
A number of studies explicitly consider the interplay between recommender systems, network structure, and social dynamics. 
\citet{cinus2022effect} combine Monte Carlo simulations, people recommenders, and opinion-dynamics models to study echo chambers and polarization under different levels of homophily and modularity. \citet{fabbri2022exposure} and  \citet{ferrara2022link} simulate repeated interactions between users and link recommenders, showing that people recommenders can generate long-term exposure inequalities and rich-get-richer effects at both group and individual levels. 
Other works study algorithmic bias and opinion dynamics under different network structures, focusing on consensus, polarization, echo chambers, or minority visibility \cite{pansanella2022mean,pansanella2022modeling,peralta2021effect,peralta2021opinion,perra2019modeling,sirbu2019algorithmic,valensise2023drivers}. 
These studies are close to ours in their use of simulation and network structure, but they mainly focus on opinion dynamics or link recommendation rather than on feed creation and {the visibility of {content} and creators}.

{\paragraph{Position of our work.} This overview highlights a key gap: the interplay between recommender systems and social network topology in shaping visibility allocation on social media platforms remains underexplored. 
We address this gap by examining how recommendation strategies and network structure \textit{jointly} influence the visibility of content and creators.
In contrast to prior work that focuses on user-side outcomes, we adopt a system-level perspective on visibility. 
Moreover, unlike studies centered on people recommendation (i.e., suggesting a user to follow), we focus on feed generation. 
This perspective allows us to isolate how design choices redistribute attention, quantify how much of the active content catalog remains unseen, and assess whether early engagement translates into sustained exposure.}

\section{Methods}
\label{sec:method}

\subsection{Experimental Settings}

\label{subsec:experimental-setting}

We conduct simulations
using YSocial,
a social media virtual twin implementing the main structural and behavioral primitives of microblogging platforms such as X, Mastodon, and Bluesky~\cite{rossetti2026ysocial}.

YSocial consists of three components: \emph{(i)} a server that provides platform primitives -- publishing, commenting, sharing, reacting, and following -- and stores simulation data; \emph{(ii)} a client that models agent behavior and communicates with the server; and \emph{(iii)} a Large Language Model (LLM) component for content generation and agent decision-making. 
{\vp In this study, the LLM component is disabled to isolate the effects of recommendation systems and network structure under controlled conditions.}

{\vp
\paragraph{Simulation Dynamics.}
Each simulation involves a set of agents \(\mathcal V\),
connected through a static social network
\(G{=}(\mathcal V,\mathcal E)\).
At each round, agents become active according to an hourly activity
distribution fitted from public Bluesky data \citep{failla2024m}.

Each agent can perform one of three actions -- \textit{publishing}, \textit{commenting}, \textit{reacting} -- according to fixed probabilities $\theta_{\mathrm{pub}},
\theta_{\mathrm{com}},  \theta_{\mathrm{react}}$.
{\lp Because these probabilities are identical across agents, agents differ only in their network position and the content they encounter, not in their behavior.}


{\lp When an agent chooses \textit{publishing}, a post is added to the system.
A post is a unit of content identified by its creator, creation round, {\vp topic label} and placeholder text.}
Comments use the same representation as posts, with the addition of a reference to the parent post. 
In the remainder of the paper, we use the terms post and content interchangeably to refer to the union of posts and comments. {\color{black}Comments are therefore recommendable content: 
they enter the active catalog, accumulate their own popularity score, and can be returned in a feed.
} 

{\lp When an agent selects either \textit{commenting} or \textit{reacting}, the recommender system generates a feed of up to $k$ posts, ranked according to its recommendation strategy. The feed is constructed from posts created during the previous $\tau_\ell$ rounds. 
The agent then selects one post uniformly at random and either comments on it or leaves a reaction.
}

Once all agents have performed their actions, the round finishes and the simulation moves to the next round. 
The simulation evolves for \(D\) simulated days. 
Time is discretized into hourly rounds, yielding \(T=24D\) rounds. 

In our {\vp reference} experiment, we set the following parameter values:
\begin{itemize}
    \item \(N=1000\) agents;
    \item \(D=30\) days, hence \(T=720\) rounds;
    \item \(k=12\) posts;
    \item \(\tau_\ell=36\) rounds;
    \item $\theta_{\mathrm{pub}}=0.30,\theta_{\mathrm{com}}=0.50,\theta_{\mathrm{react}}=0.20$.

\item The social network $G$ is generated using the preferential attachment model \citep{barabasi1999emergence} with \(N=1000\) nodes and attachment parameter \(m=10\), {\vp and is treated as static and undirected throughout the simulation.} 
\end{itemize}

We vary the recommender system while holding the remaining settings fixed.
For each recommender configuration, we perform \(J{=}10\) independent runs to assess cross-run variability.
All remaining YSocial settings are held constant.
}

\smallskip 

{\vp  We also perform a \textbf{sensitivity analysis} to assess how these choices affect our results, as the reference parameters represent plausible values for the simulated environment rather than empirically calibrated estimates.
Firstly, we conduct a \textit{topology ablation} by replacing the BA network with Erd\H{o}s--R\'enyi (ER) \cite{erdhos1959random} and power-law-cluster (PC) \cite{holme2002growing} networks while retaining the same reference settings. 
Secondly, we vary \textit{one factor at a time (OFAT)} across all three topologies, considering simulation duration $D\in\{30,60,90\}$ days, population size $N\in\{1000,5000,10000\}$, visibility window $\tau_\ell\in\{24,36,48\}$ hours, and feed size $k\in\{6,12,24\}$. {\color{black} When \(N\) varies, the network-generation parameters are adjusted so that network density remains approximately constant; this experiment therefore measures sensitivity to \emph{system scale} rather than to population size in isolation.}
For each topology--parameter combination, we perform \mbox{\(J = 5\)} independent simulation runs, except for the reference configuration, which uses  \mbox{\(J = 10\)} runs.}

\subsubsection{Recommender Systems.}
\label{sec:rec_strategies}
We implement six recommender systems that isolate common feed-generation logics. 
Three global recommender systems rank content from the whole active content catalog. 
Three network-aware recommender systems use the social network as part of candidate filtering, prioritizing content produced by followed accounts before applying the corresponding ranking rule.

\paragraph{\textit{Global recommender systems}\newline}

\noindent\textit{\underline{Reverse-Chronological (RC)}}
ranks currently visible contents by creation time in descending order. It serves as the baseline for global recommender systems and represents a minimal ordering rule without any personalization, popularity weighting, or network awareness.

\noindent\textit{\underline{Popularity (P)}}
{\lp ranks visible content by its popularity, measured as a weighted combination of user engagement features. Specifically, each reaction contributes two points and each comment one point, using the interactions cumulated by the content up to the moment the feed is
generated. {\lp Throughout the paper, references to reactions in the context of popularity denote this compound score.} Ties are broken by reverse chronological order.}


\noindent\textit{\underline{User--User Collaborative Filtering (UCF)}}
identifies users with similar liking histories and recommends contents liked by these similar users but not yet seen by the target user. Candidates are ranked by the number of similar users who liked them, with recency as a tie-breaker. {\vp Implementation details are reported in Appendix~\ref{app:recommender-details}.}

\paragraph{\textit{Network-aware recommenders}\newline}
\noindent\textit{\underline{Follower-Reverse-Chronological (F)}}
allocates up to 60\% of the timeline to contents from followed creators, ranked in reverse-chronological order, and fills the remaining slots from the global catalog in the same order.
It is the baseline for network-aware recommender systems.

\noindent\textit{\underline{Follower-Popularity (FP)}}
uses the same follower-to-global slot allocation. Within both pools, contents are ranked by {\lp their popularity}.

\noindent\textit{\underline{Linear Ranker (LR)}}
combines reverse-chronological follower, global popularity, and user--user collaborative filtering. 
Candidate contents are reranked through a weighted linear score based on recency, followed-author status, user--author affinity, topic similarity, and similar-user signals\footnote{\url{https://github.com/twitter/the-algorithm}}.
Implementation details are reported in Appendix~\ref{app:recommender-details}.

\subsection{Metrics}
\label{subsec:metrics}

{\vp
We evaluate visibility along three dimensions:
\textit{(i)} {\upd concentration} and catalog coverage,
\textit{(ii)} the relationship between creators' network position and visibility, and
\textit{(iii)} the association between accumulated popularity and subsequent visibility.
Recommendation volume provides the underlying measure of visibility at both the content and creator levels.
All metrics are computed separately for each experimental condition.

{\lp For each recommender system $f$, we compute the mean value $\overline{M}^{(f)}$ of each metric across simulation runs. 
When applicable, we also report the difference $\Delta \overline{M}^{(f)}$ with respect to the corresponding baseline, where the baseline is RC for global recommenders and F for network-aware recommenders. {\color{black} A direct comparison between the two baselines, which shows that they are not equivalent and motivates treating them separately, is reported in Appendix~\ref{sec:appendix-comparison}.}


}

\paragraph{Recommendation volume.}
{\lp 
{\vp Recommendation volume} quantifies how many times content and creators appear in users' feeds. 
We denote by $r_p$ the recommendation volume of a content item $p$, i.e., the total number of times it appears in users' feeds throughout a simulation. Similarly, $r_a$ denotes the recommendation volume of a creator $a$, computed as the total recommendation volume of all content created by $a$. 
}



{\lp
To understand how recommendations are distributed across content and creators, we estimate the empirical
{\vp probability mass function (PMF)}
of recommendation volumes.

For $x \in \{p,a\}$, we denote by
{\vp $\phi_x^{(f,j)}(r)$ the fraction of entities of type $x$ that receive exactly $r$ recommendations in run $j$ under recommender system $f$, considering only entities that receive at least one recommendation ($r>0$). By construction, each PMF is normalized over this set.
To aggregate across runs, we align the run-level PMFs on the union of
the observed supports and average them with equal weight per run.}

}


\paragraph{Concentration.}

{\lp 
It measures how unevenly recommendation volume is allocated across content or creators. We quantify it using the Gini coefficient, a standard measure of inequality widely used to assess aggregate diversity in recommender systems \cite{vargas2014improving}. 

Let $r_{x,(1)} \leq \dots \leq r_{x,(n_x)}$ denote the recommendation volumes of entities of type $x\in \{p,a\}$, sorted in non-decreasing order, where $n_x$ is the total number of entities. {\vp The catalog includes entities that receive zero recommendations.}
The run-level Gini coefficient is:
\begin{equation}
    G_x
    =
    \frac{
        \displaystyle
        \sum_{i=1}^{n_x}
        \left(2i-n_x-1\right)r_{x,(i)}
    }{
        \displaystyle
        n_x\sum_{i=1}^{n_x}r_{x,(i)}
    }.
    \label{eq:gini}
\end{equation}
Larger values indicate that recommendation volume is concentrated on a smaller subset of content or creators. 

We average $G_x$ across simulation runs and compare the resulting value with the matched baseline. Positive values of $\Delta \overline{G}_x^{(f)}$ indicate greater concentration than the baseline, whereas negative values indicate a more even allocation. {\vp Throughout the paper, we use the terms  \emph{concentration}, and \emph{inequality} interchangeably to refer to this quantity.}
}
\paragraph{Coverage.}
\label{subsubsec:diversity}

{\lp 
This metric quantifies the fraction of the content or creator catalog that receives at least one recommendation. 
It follows the notion of catalog coverage in recommender-system evaluation \cite{herlocker2004evaluating}. 

For $x\in\{p,a\}$, we denote the run-level coverage by $C_x$, computed as the share of entities with recommendation volume $r_x>0$. 
Coverage ranges from {\upd 0 to 100\%}, with larger values indicating that recommendation exposure reaches a larger portion of the catalog. 

We average $C_x$ across runs and report its difference from the matched baseline. Positive values of $\Delta \overline{C}_x^{(f)}$ indicate broader catalog coverage than the baseline, whereas negative values indicate that a larger share of the catalog remains unexposed.
}


\paragraph{Creator visibility by degree.}

{\lp 
To assess whether recommendation systems amplify structural advantage, we examine how creator visibility varies with degree in the social network. 
We consider two complementary measures: recommendation share and unique reach.

{\upd Within each simulation run, creators are ranked by their network degree $k_a$ and grouped into $Q_h$, the $h$-th of 20 equally spaced degree-percentile intervals (ventiles).} For each creator $a$, the recommendation share $s_a$ is the fraction of all recommendations received by that creator, while the unique reach $u_a$ is the fraction of users exposed at least once to one of the creator's posts. 
For each {\upd percentile interval $Q_h$}, we compute the mean recommendation share {\upd $s(Q_h)$} and the mean unique reach {\upd $u(Q_h)$} across the creators assigned to that {\upd interval}.

We average these quantities across simulation runs and report {\upd $\Delta \overline{s}^{(f)}(Q_h)$ and $\Delta \overline{u}^{(f)}(Q_h)$} relative to the matched baseline. 
Positive values indicate that recommendation strategy $f$ allocates a larger recommendation share or reaches a larger fraction of users than the baseline for creators in {\upd percentile interval $Q_h$}. 
Runs that contain no creators in a given {\upd interval} do not contribute to that {\upd interval}'s average.

{\upd Ranking within each run makes one interval mark the same structural position in every topology, whereas the absolute degree does not. Creators of equal degree share a rank and are never split across intervals, so intervals may differ in size or be empty; an empty interval is reported as missing rather than as zero.
}
}

\paragraph{Popularity--future-exposure correlation.}

{\lp 
{\upd This metric assesses whether the popularity a post has accumulated at a given point in its lifecycle predicts the recommendation exposure it receives afterwards. Rather than fixing a single early window, we evaluate this association at every content age.
We restrict the analysis to posts whose full eligibility window falls within the simulation, since posts published near the end would have their future exposure truncated.}

For a post $p$ and content age $t_p$, we define early popularity $\pi_{p, t_p}$ as the cumulative popularity score observed before age $t_p$, and subsequent exposure $z_{p,t_p}$ as the total number of recommendations received from age $t_p$ until the end of the post's eligibility window. 
We then compute the Spearman correlation $\rho_{t_p} =
\operatorname{corr}_{\mathrm{S},\,
        p\in\mathcal P_{\tau_\ell}}
    \left(
        \pi_{p,t_p},
        z_{p,t_p}
    \right)$
across eligible posts.
We average $\rho_{t_p}$ across simulation runs. 
Higher positive values indicate that posts with greater early popularity tend to receive more recommendation exposure during the remainder of their lifecycle.
}

\medskip

For the sensitivity analysis, for each configuration we compute the same metrics described above as in the reference experiment, and report their variation relative to the reference configuration; {\vp for clearer comparison, } popularity reinforcement is summarized by the interval-normalized area under the popularity--future-exposure curve, \(\mathrm{AUC}_\rho\).

\section{Results}
\label{sec:results}
{\upd We present our results in four parts. We first analyze global
recommenders in terms of recommendation-volume distributions, concentration,
and coverage, separately for contents and creators. We then examine
network-aware recommenders using the same measures, complemented by
degree-resolved recommendation share and unique reach. We next turn to
popularity reinforcement, which we analyze across both recommender classes.
Finally, we assess the robustness of these findings through two sensitivity
analyses: a topology ablation and a one-factor-at-a-time (OFAT) analysis, the
latter organized parameter by parameter. Detailed results are reported in
Appendices~\ref{sec:appendix-network-ablation} and~\ref{app:ofat},
respectively.}

The first three parts refer to the reference setting described
in Experimental Settings. The magnitudes we report are therefore properties
of that setting; the Sensitivity Analysis shows how they vary across
topologies and parameters, and which comparisons between recommenders
hold.
}
\begin{figure}[t!]
\centering
\includegraphics[width=\linewidth]{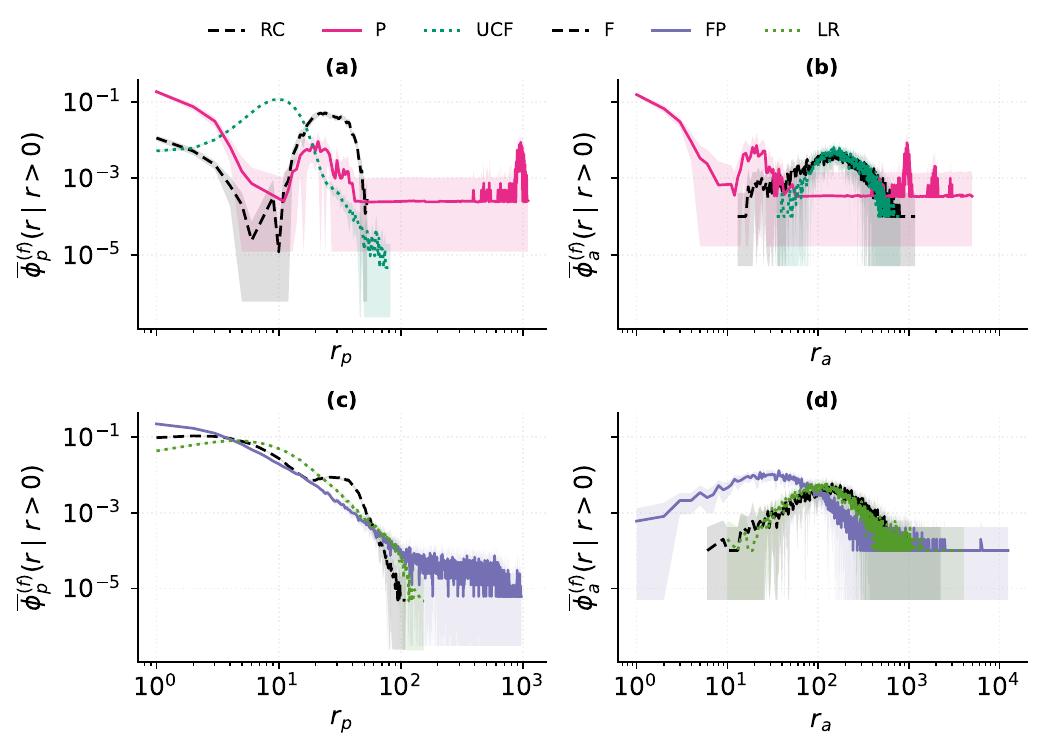}
\caption{
  \textbf{Recommendation-volume distributions in the BA topology.}
  Panels~(a) and~(b) report the mean probability mass functions
  \(\overline{\phi}_p^{(f)}(r\mid r>0)\) of content-level recommendation
  volume \(r_p\) and \(\overline{\phi}_a^{(f)}(r\mid r>0)\) of creator-level
  recommendation volume \(r_a\) for the global recommenders (RC, P, UCF).
  Panels~(c) and~(d) report the same two distributions for the
  network-aware recommenders (F, FP, LR).
  Shaded regions indicate cross-run variability; both axes are shown on
  logarithmic scales.
}
\label{fig:ba-volume}
\end{figure}
\subsection{Global recommender systems}
Global recommender systems operate independently of the {social} network, isolating the contribution of ranking {strategy} to visibility allocation before examining how network structure modulates these effects.
We use RC as the baseline and compare it with P and UCF.
\begin{mdframed}[backgroundcolor=gray!5, linewidth=0.5pt]
\small \textbf{Takeaway:}  Without network filtering, the ranking rule alone
sets the visibility regime. Differently from the reverse chronological baseline, ranking by popularity creates a winner-selection mechanism: it concentrates exposure and reduces coverage, for both content and creators. Collaborative filtering does the opposite, keeping nearly all active content and every creator in circulation.
\end{mdframed}

\paragraph{Contents.}
{\lp The popularity-based recommender (P) produces the most polarized visibility distribution (Figure~\ref{fig:ba-volume}a, pink line). Most content receives a few recommendations, while a small subset accumulates high visibility. As a result, P largely suppresses moderate levels of exposure, assigning substantially less probability mass than RC to recommendation volumes between 10 and 100, while producing a long right tail extending to about 1,000 recommendations.

UCF exhibits the opposite pattern (Figure~\ref{fig:ba-volume}a, green line). Recommendation volume is distributed more evenly across the catalog, with greater probability mass at low and intermediate exposure levels and a much less pronounced right tail. Rather than concentrating visibility on a few highly exposed items, UCF spreads moderate visibility across a larger fraction of content.}

\begin{figure}[t!]
\centering
\includegraphics[width=\linewidth]{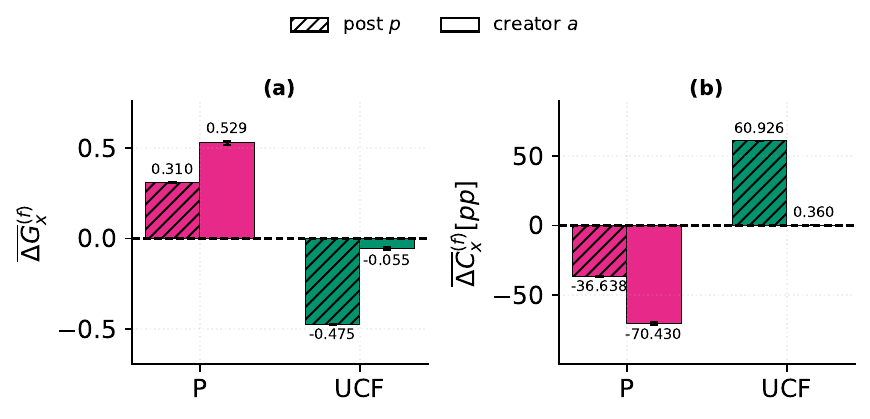}
\caption{%
  \textbf{Concentration and coverage under global recommenders (BA topology).}
  Panel~(a) reports changes in concentration  
  {\vp(\(\Delta\overline{G}_x^{(f)}\))}
  and panel~(b) reports changes in coverage 
  {\vp(\(\Delta\overline{C}_x^{(f)}\))}, both relative to RC.
  Hatched bars denote contents; plain bars denote creators.
  Error bars indicate cross-run standard deviation.
}
\label{fig:ba-global-regime}
\end{figure}

{\lp Figure~\ref{fig:ba-global-regime} confirms this pattern. 
The RC baseline already exhibits moderate content-level inequality ($G_p=0.68$). 
Popularity amplifies this effect, pushing inequality close to its maximum ($G_p=0.99$, $+0.31$) while reducing content coverage to 1.8\% ($-36.6$~pp): fewer than 2 in 100 content items receive any recommendation, and the top 10\% capture almost all exposure. 
In contrast, UCF largely eliminates this concentration ($G_p=0.20$, $-0.48$) and increases coverage to 99.4\% ($+60.9$~pp), i.e., nearly all active content receives at least one recommendation. 
}


\paragraph{Creators.}
{\lp Compared with RC, P produces a much more unequal creator-visibility distribution (Figure~\ref{fig:ba-volume}b, pink line). Most creators receive little visibility, while a minority accumulates orders of magnitude more recommendations, yielding a long right tail. 
In contrast, RC and UCF produce unimodal distributions centered at intermediate recommendation volumes, indicating a more balanced allocation of visibility across creators.}


{\lp P increases creator-level inequality even more than content-level inequality, raising the Gini coefficient to $G_a=0.85$ ($+0.53$) and reducing creator coverage to 29.2\% ($-70.4$~pp). 
As a result, more than two in three creators receive no recommendation at all. In contrast, UCF leaves creator-level inequality close to the RC baseline ($G_a=0.26$) while maintaining full creator coverage. 
This complete coverage arises because UCF generates personalized recommendation lists. Since similar-user neighborhoods differ across users, recommendations are distributed across a broader set of content and, consequently, creators.}

\begin{figure}[hbpt]
    \centering
    \includegraphics[width=\columnwidth]{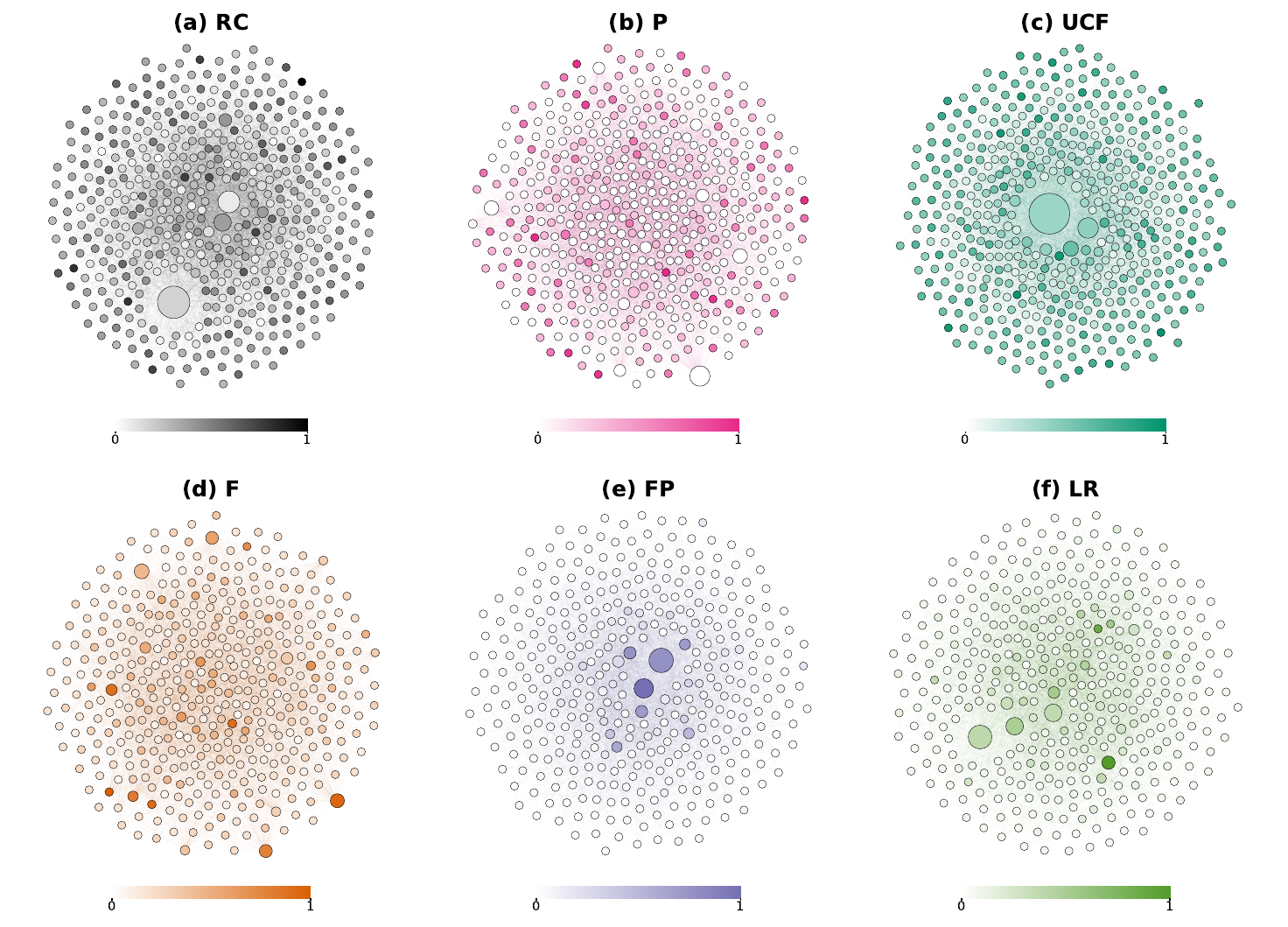}
\caption{%
  \textbf{Creator visibility for each recommender system (BA topology).}
  Each panel shows creator visibility in the BA social network under a 
  different feed: (a)~RC, (b)~P, (c)~UCF, (d)~F, (e)~FP, (f)~LR. 
  Node size is proportional to degree; node saturation reflects normalized 
  visibility. {\vp Each panel shows the union of the 300 nodes with the highest degree and the 300 creators with the highest recommendation exposure.}
}
    \label{fig:ba-network-viz}
\end{figure}

{\lp Figure~\ref{fig:ba-network-viz}a-c illustrates the different visibility regimes. 
Nodes are sized by degree and colored by the total recommendation volume received by each creator. RC produces relatively uniform exposure across the network, whereas P concentrates visibility on a handful of creators, yielding a winner-takes-all pattern largely independent of degree. UCF instead distributes exposure much more evenly, in line with its near-complete creator coverage.}



\subsection{Network-aware recommender systems}
{\lp We next examine network-aware recommendation strategies, which restrict the candidate pool using the follower graph. This allows us to assess how network structure and ranking jointly shape visibility allocation. Here, F serves as the baseline, while FP adds popularity ranking and LR combines multiple signals through linear reranking.}


\begin{mdframed}[backgroundcolor=gray!5, linewidth=0.5pt]
\small \textbf{Takeaway:} When the follower graph determines candidate
selection, ranking and network structure interact. 
In this case, ranking by popularity keeps more content circulating than the corresponding global ranking, but ties creator visibility to network position: high-degree creators gain  visibility and reach, low-degree creators gain little or nothing. Multi-signal ranking favors high-degree creators too, but far more weakly, and never raises reach above the follower baseline.
\end{mdframed}


\begin{figure}[b!]
\centering
\includegraphics[width=\linewidth]{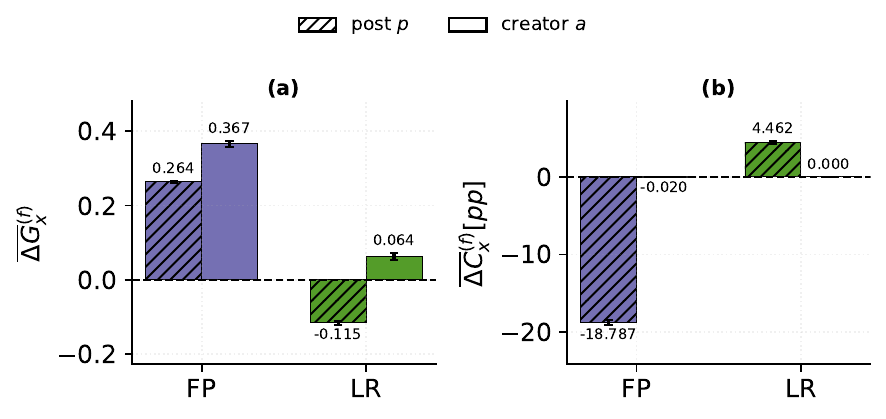}
\caption{%
  \textbf{Concentration and coverage under network-aware recommenders (BA topology).}
  Panel~(a) reports changes in concentration {\vp ($\overline{\Delta G^{(f)}_x}$)}
  and panel~(b) reports changes in coverage {\vp ($\overline{\Delta C^{(f)}_x}$)},
  both relative to F.
  Hatched bars denote contents; plain bars denote creators.
  Error bars indicate cross-run standard deviation.
}
\label{fig:ba-network-regime}
\end{figure}

\paragraph{Contents.}
{\lp Figure~\ref{fig:ba-volume}c--d shows the content- and creator-level recommendation-volume distributions, while Figure~\ref{fig:ba-network-regime} summarizes the corresponding inequality and coverage metrics. 
The follower-based baseline (F) already achieves high content coverage ($C_p=93.1$\%), substantially exceeding the global RC baseline (38.5\%). 
This reflects the effect of restricting competition to local follower neighborhoods, allowing content that would quickly disappear in a global feed to remain visible.

Starting from this broad-coverage regime, FP substantially increases concentration, raising the content-level Gini coefficient to $G_p=0.84$ ($+0.27$) while reducing coverage to 74.3\% ($-18.8$~pp). 
In contrast, LR further broadens content circulation, lowering inequality ($G_p=0.46$, $-0.11$) and increasing coverage to 97.6\% ($+4.5$~pp), comparable to UCF. 
Overall, coupling the follower graph with popularity reproduces the concentration pattern observed under P, whereas multi-signal reranking preserves -- and even extends -- the broad circulation of the follower baseline.}


\paragraph{Creators.}
{\lp The creator-level picture differs markedly from the content level. 
As shown in Figure~\ref{fig:ba-volume}d, FP spreads creator visibility over a wider range, with a pronounced right tail extending to much higher recommendation volumes. 
This translates into a substantial increase in creator-level inequality ($G_a=0.73$, $+0.36$), approaching the value observed under P ($G_a=0.85$), despite FP operating on a follower-filtered rather than global candidate pool. 
Notably, the increase in creator-level inequality is larger than the corresponding increase at the content level, indicating that FP concentrates visibility on a relatively small set of creators. 
In contrast, LR largely preserves creator-level equality: although it slightly increases $G_a$ relative to the follower baseline, the effect is much weaker than under FP. 
Creator coverage remains close to 100\% for all three recommendation systems, indicating that the follower graph ensures that nearly every active creator reaches at least one user, while the ranking strategy primarily determines how much visibility each creator receives.}


\begin{figure}[ht]
\centering
\includegraphics[width=\linewidth]{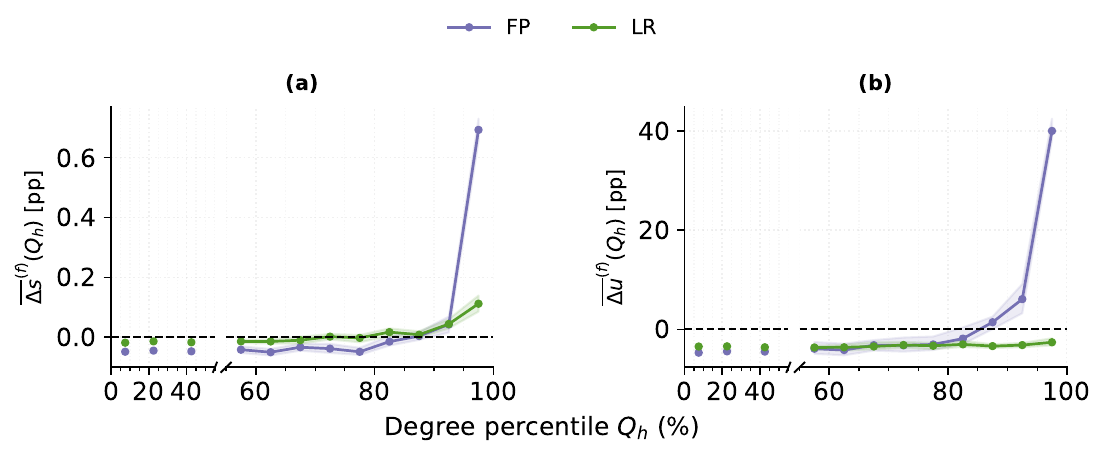}
\caption{%
  \textbf{Degree-resolved creator visibility under network-aware recommenders (BA topology).}
  Panel~(a) reports changes in mean recommendation share {\vp ($\overline{\Delta s}^{(f)}(Q_h)$)}
  and panel~(b) reports changes in mean unique reach {\vp ($\overline{\Delta u}^{(f)}(Q_h)$)},
  both relative to F and expressed in percentage points.
  {\upd Creators are placed in 20 tie-preserving percentile intervals $Q_h$; the axis breaks at the 55th percentile to compress the sparse peripheral intervals, and is linear within each segment.}
  Shaded areas indicate cross-run variability.
}
\label{fig:ba-degree-amplification}
\end{figure}


{\lp The degree-resolved analysis in Figure~\ref{fig:ba-degree-amplification} 
{\upd shows the degree dependence underlying FP's creator-level concentration.}
Under FP,
{\upd the differences in both mean recommendation share and unique reach relative to F become increasingly positive with creator degree, with the strongest amplification concentrated among the highest-degree creators.}
Relative to F, 
{\upd low- and intermediate-degree creators generally receive lower recommendation share and unique reach, whereas the creators in the top degree ventile benefit substantially, with unique-reach gains reaching about 40 percentage points.}
{\color{black} LR 
{\upd shows a much weaker degree gradient in recommendation share, while unique reach remains consistently below F and varies little with creator degree.}}}

This reveals a key interaction between ranking strategy and network structure: when popularity ranking operates over the follower graph, it amplifies the visibility advantage of highly connected creators. Importantly, this degree gradient is absent under global popularity (P; Appendix~\ref{sec:appendix-global-degree}, Figure~\ref{fig:appendix-global-degree-ba-er}), indicating that it emerges from the interaction between popularity ranking and network structure rather than from popularity alone.

The network visualization in Figure~\ref{fig:ba-network-viz}d--f illustrates the same pattern. 

\subsection{Popularity reinforcement}

{\lp Finally, we examine whether early popularity predicts subsequent recommendation visibility and how this relationship varies across recommendation strategies. 
We measure early popularity as the reactions accumulated up to content age $t_p$ within the 36-hour lifecycle. 
We focus on content-level analysis, as aggregating reactions across posts would confound popularity with differences in creator activity.}


\begin{mdframed}[backgroundcolor=gray!5, linewidth=0.5pt]
\small \textbf{Takeaway:} The coupling between early engagement and later exposure is determined by the ranking rule. This coupling is nearly complete under global popularity ranking, weaker when popularity is used to rank only posts by followers, and essentially absent under collaborative filtering.
\end{mdframed}

\paragraph{Contents.}

\begin{figure}[t!]
\centering
\includegraphics[width=\linewidth]{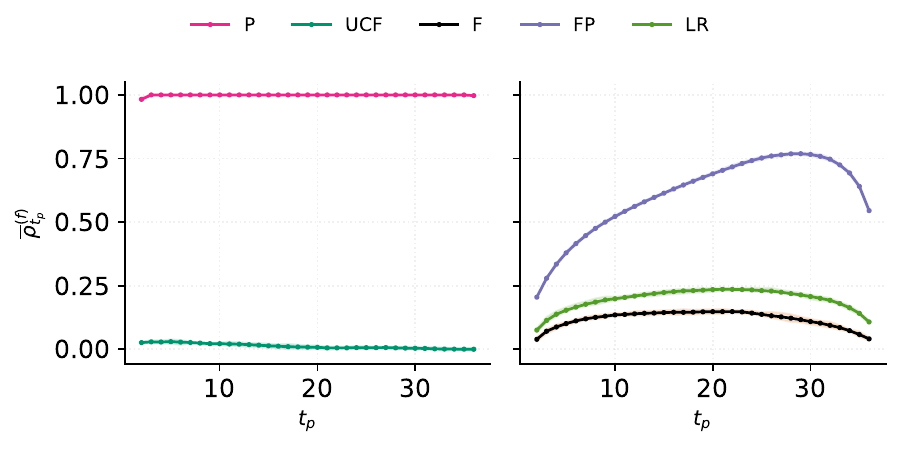}
\caption{%
  \textbf{Popularity reinforcement (BA topology).}
  Spearman correlation {\vp $\overline{\rho}_{t_p}^{(f)}$} between early popularity
  (reactions accumulated up to content age $t_p$)
  and future recommendation exposure after $t_p$.
  Panel~(a) shows global recommenders (P and UCF);
  RC is omitted because its correlation is defined over at most the first two content ages
  and provides no meaningful trajectory.
  Panel~(b) shows network-aware recommenders (F, FP, and LR).
  Values are reported in absolute terms;
  shaded areas indicate cross-run variability.
}
\label{fig:ba-popularity-reinforcement}
\end{figure}

{\lp Figure~\ref{fig:ba-popularity-reinforcement} shows that popularity reinforcement strongly depends on the ranking strategy. Among global recommenders (panel a), P and UCF represent opposite extremes. 
Under P, the correlation between early popularity and subsequent exposure rapidly approaches one ($\rho_{t_p}=0.98$ by $t_p=2$), indicating that content visibility is largely determined within the first hours of its lifecycle. Under UCF, the correlation remains close to zero at all ages ($\rho_{t_p}<0.03$), showing no systematic coupling between early reactions and later exposure.
RC is excluded from the plot because recency ranking leaves virtually no future exposure for older content, preventing popularity reinforcement from developing. In other words, under RC, the recommendation window closes before early reactions can systematically translate into later exposure.}


{\lp Among network-aware recommenders (Figure~\ref{fig:ba-popularity-reinforcement}b), all three strategies show a bell-shaped trajectory, but differ markedly in the strength of popularity reinforcement. 
FP exhibits the strongest coupling, with $\rho_{t_p}$ peaking at approximately $0.77$, substantially below global P but well above F and LR. 
In contrast, reinforcement remains weak under F ($\rho_{t_p}<0.15$) and LR ($\rho_{t_p}<0.24$) throughout the content lifecycle.}



{\lp These patterns mirror the visibility distributions: P and FP, which produce the most concentrated and lowest-coverage content visibility, also exhibit the strongest coupling between early engagement and later exposure. 
This coupling is weaker under FP than P (peak $\rho \approx 0.77$ vs. $\rho \rightarrow 1$), suggesting that the follower graph moderates popularity reinforcement by providing alternative pathways to visibility. {\color{black} Since reactions are random, cumulative advantage under P follows largely from
the ranking rule itself, which feeds accumulated engagement back into rank.
What the rule does not fix is its strength: comparing P with FP, F and LR shows
how much of the coupling survives within follower-based pools or alongside
other signals.}}



{\vp
\subsection{Sensitivity Analysis}
\label{subsec:sensitivity-analysis}
We assess robustness {\lp of our results} along two dimensions: the topology of the social {\lp network}, and
the values of the main simulation parameters, as described in Methods. Full results discussion is reported in
Appendix~\ref{sec:appendix-network-ablation} and Appendix~\ref{app:ofat};
additional figures and interactive comparisons are available in the online
dashboard.\footnote{\url{https://ofat-sensitivity-dashboard.github.io/}}

\begin{mdframed}[backgroundcolor=gray!5, linewidth=0.5pt]
\small \textbf{Takeaway.} Recommender-specific visibility patterns are preserved across sensitivity settings, but their magnitude is configuration-dependent. 
The random topology distributes creator visibility more evenly across degree classes;
larger feeds broaden exposure and reduce concentration; longer content lifetimes reduce coverage and strengthen popularity-driven concentration; system scale changes the relative behavior of recommendation mechanisms by increasing competition for exposure and enlarging the neighborhoods over which network-aware feeds operate.
\end{mdframed}

\paragraph{{\lp Network} topology ablation.} 
Changing the topology {\lp of the social network} has no effect on visibility under global ranking strategies, but produces small differences for network-aware feeds. 
The main difference is between random and scale-free topologies: degree
heterogeneity produces more unequal exposure, with the strongest differences
under popularity ranking.
{\lp This happens} because high-degree nodes receive disproportionate exposure amplification, whereas in random networks visibility is distributed more evenly across degree classes. 
Introducing additional clustering in scale-free networks makes no substantial difference, suggesting that degree heterogeneity
contributes more strongly than local clustering to visibility patterns. Under FP,
however, popularity reinforcement is stronger in the random topology even though
visibility inequalities are weaker. 
Overall, {\lp network} topology does not substantially
change the visibility regime induced by a recommender {\lp system}, but changes the strength
and source of visibility inequalities within that regime.

\paragraph{Content visibility window.}
Content lifetime primarily affects visibility under popularity-based
recommenders. At the content level, longer visibility windows reduce
coverage and allow popularity reinforcement to persist for longer under FP, while
content-level concentration and the recommendation-volume distributions change
comparatively little; RC, UCF, F, and LR are largely insensitive. At the
\textbf{creator} level, longer windows increase concentration and reduce coverage
most clearly under P, while the degree-resolved patterns remain stable. The
visibility window therefore primarily affects how long popularity advantages can
persist, and does not substantially alter how visibility varies with creator
degree.

\paragraph{Simulation duration.}
Simulation duration has almost no effect on content-level visibility. At the creator level, a longer simulation reduces the number of creators receiving no recommendations, particularly under global ranking strategies. 
This reduction is strong enough for P to become less concentrated than FP under scale-free topologies, showing that the relative concentration of global and follower-based popularity ranking depends on the simulation horizon.

\paragraph{Feed size.}
Feed size controls the amount of exposure available for allocation. At the content level, increasing \(k\) increases coverage and reduces concentration across recommender {\lp systems} and topologies. {\lp This happens} most markedly under RC because concentration under recency ranking depends largely on the number of available feed slots relative to the size of the active catalog. 
Visibility remains highly unequal under P and comparatively evenly distributed under UCF throughout the tested range. 
The effect on popularity reinforcement is almost negligible, except
for FP, where it decreases as the feed grows. At the creator level,
larger feeds reduce concentration and increase coverage, while the
degree-dependent visibility patterns remain essentially unchanged: additional
feed slots reduce competition for visibility without altering the relative
advantage of high-degree creators under FP and LR.

{\upd \paragraph{Population size.} 
Population size produces the strongest sensitivity at the creator level and also
affects content visibility for several recommender {\lp systems}. Because increasing \(N\)
also expands the content catalog and, under the approximately fixed-density
scaling used here, increases the average degree, we interpret this experiment as
sensitivity to \emph{system scale} rather than as the isolated effect of
population size. At the content level, increasing scale generally
increases concentration and reduces coverage under RC, P, F, FP, partly because
the catalog grows relative to a fixed number of recommendation slots. More
importantly, FP becomes progressively more selective and approaches the
visibility regime of global P, its popularity reinforcement becoming nearly
maximal at the largest tested populations. 
At the creator level, it is the only network-aware recommender whose near-complete coverage does not persist.
Global rankings therefore become increasingly constrained by competition for
exposure, while the combination of expanding follower neighborhoods and
popularity ranking makes FP substantially more selective at larger scales.}
\section{Discussion}
\label{sec:discussion}
{\lp We have shown how visibility in social-media feeds is shaped not by recommender systems or network structure alone, but by their interaction. 
As a stylized mechanism study, we used a controlled simulation to compare six prototypical recommendation strategies based on recency, popularity, collaborative filtering, follower-based filtering, and their combinations. 
The limited variability across independent runs suggests that the observed patterns are systematic properties of the recommender--topology combinations considered, rather than artifacts of simulation noise. 
{\upd The main findings persist when the simulation parameters are varied, while the magnitude of visibility concentration and coverage, and some relative comparisons between recommenders, depend on the simulation settings.}} 

{\lp A first main finding is that the \textbf{recommendation strategy is the primary determinant of the visibility regime}. Popularity produces the most unequal content-level outcome by repeatedly directing exposure toward a restricted subset of contents, leaving much of the active catalog unseen. Collaborative filtering exhibits the opposite pattern, keeping nearly the entire active catalog in circulation while reducing concentration.}
This contrast with studies showing that collaborative filtering amplifies popularity bias when trained on long-tailed interaction logs \cite{abdollahpouri2020multi} should be interpreted in light of our stylized setting. In our simulations, reactions are random, content semantics are absent, and the active catalog is continuously renewed, so collaborative filtering does not inherit a stable popularity structure. Accordingly, our findings isolate the effect of similarity-based allocation rather than characterize the behavior of deployed collaborative filtering systems. They also reinforce the multi-stakeholder perspective that coverage alone is insufficient, as visibility should be evaluated at the creator level and in terms of exposure volume \cite{abdollahpouri2020multi}. 
{\upd These findings remain visible across parameter variation, although their magnitude depends on the amount of exposure available and on the simulation duration.}} 

{\lp A second main finding is that \textbf{network-aware recommendation couples visibility to creators' structural position}. The follower graph broadens content circulation relative to the corresponding global strategy because content competes within local neighborhoods rather than a single platform-wide pool. At the same time, this structure creates unequal opportunities for exposure: high-degree creators can enter a larger number of follower-based candidate pools and therefore receive disproportionate visibility. 
These findings qualify the view of network-aware recommendation as inherently diversity-enhancing: it can broaden content circulation, yet still concentrate visibility on highly connected creators. This is consistent with empirical work showing that exposure reflects the joint effects of network structure, algorithmic filtering, and user choice \cite{bakshy2015exposure}, as well as evidence that attention inequality can emerge from the social network itself \cite{zhu2016attention}, for example through people recommenders \cite{fabbri2022exposure}.} 
{\upd Under homogeneous degree distributions, creator visibility is more evenly distributed across degree classes, whereas degree-heterogeneous networks amplify high-degree creators. 
At larger scales, fixed feed capacity interacts with greater content availability, amplifying the concentration induced by popularity-based selection.
Thus, the visibility effects of network-aware recommendation depend on both network structure and system scale.}

A third main finding concerns \textbf{popularity reinforcement as a temporal visibility mechanism}. Popularity ranking does not merely produce a concentrated allocation at a given point in time: by converting early engagement into subsequent exposure, it creates a feedback loop in which already-successful content becomes increasingly likely to receive further recommendations. The strength of this reinforcement depends on how broadly popularity is applied and on the structure of the available content. Global popularity ranking generally produces the strongest reinforcement, whereas restricting popularity ranking to follower-based content can attenuate this effect. This result is consistent with work on algorithmic amplification \cite{huszar2022algorithmic} and relates to the ``few-get-richer'' effect of popularity-based recommender systems \cite{germano2019few}.
{\upd This distinction narrows with system scale, as larger follower neighborhoods make local popularity increasingly representative of the broader system. Reinforcement is also shaped by content lifetime and feed size: longer lifetimes preserve early popularity advantages, whereas larger feeds spread exposure across more content. Network structure further modulates the process, with more homogeneous degree distributions supporting stronger temporal reinforcement and degree heterogeneity tending to localize exposure. Overall, popularity reinforcement reflects the interaction between ranking, content availability, and network structure and scale.}

\paragraph{Limitations and Future Work.} {\lp Because our goal is to isolate underlying mechanisms rather than reproduce a deployed platform,} the simulation relies on simplified user behavior. This choice excludes factors present in real systems, such as positional bias, cognitive biases, and content relevance \cite{chen2023bias}, {\upd as well as strategic adaptation by users and creators, such as optimizing posting behavior, seeking engagement, or adapting to the recommender}. Future work should replace uniform selection with attention-sensitive interaction rules calibrated on empirical data, or with LLM-enhanced agents \cite{rossetti2026ysocial}. For example, recent models of online attention could help represent how limited attention shapes interaction with dynamic feeds \cite{ojer2025modeling}. 

Content semantics {\upd coupled with heterogeneous user preferences and preference-driven selection and reactions} should also be introduced to distinguish catalog coverage from semantic or ideological diversity, since broad coverage may coexist with belief-aligned exposure \cite{nguyen2014exploring}. 
{\upd The three network models allow us to compare degree distributions and, through the power-law-cluster network, additional local clustering, but omit community structure, homophily, directed ties, and network evolution. These structural properties could alter how network position interacts with recommendation and may therefore produce visibility patterns not captured by the present ablation.} For example, homophily can reduce minority visibility under people recommendation \cite{fabbri2020effect}. Future work should therefore use networks with tunable communities or estimated from platform data, and allow the follower graph to evolve. 

The simulation also omits visibility pathways beyond the algorithmic timeline, such as shares, likes from followed accounts, or discovery outside the recommended feed. These mechanisms can substantially affect exposure \cite{guess2023reshares} and remain a challenge for recommender-system simulations \cite{chaney2021recommendation}. Incorporating them would make the framework more representative of deployed platforms while preserving its main advantage: observing both shown and unseen eligible content under controlled conditions. 

{\upd The robustness analysis is itself bounded because it tests which mechanism-level patterns persist and how their magnitude and relative expression change across the tested configurations. Future work could extend this analysis by jointly varying behavioral, structural, and ranking parameters, thereby testing interactions that OFAT cannot capture.}

\bibliography{references}

\appendix
\setcounter{secnumdepth}{1}
\section{Recommender-System Implementation Details}
\label{app:recommender-details}

All experiments were executed using the SQL backend of YSocial. The complete implementation is available in the linked code repository\footnote{\url{https://github.com/YSocialTwin/YSimulator/blob/main/YSimulator/YServer/recsys/content_recsys_db.py}}; the six
systems used in this study correspond to the modes
\textit{ReverseChrono}~(RC), \textit{ReverseChronoPopularity}~(P),
\textit{CollaborativeUserUser}~(UCF), \textit{ReverseChronoFollowers}~(F),
\textit{ReverseChronoFollowersPopularity}~(FP), and
\textit{HybridLinearRanker}~(LR). The
following descriptions therefore refer to the implementations of
User--User Collaborative Filtering and the Linear Ranker recommenders.

\paragraph{User--User Collaborative Filtering (UCF).}
The collaborative-filtering recommender uses only interactions recorded
as likes. Let \(L_u\) denote the set of contents liked by user \(u\).
Comments are stored as content items in their own right, so a like
assigned to a comment enters \(L_u\) as a separate entry and is not
attributed to the parent post.

The similarity between two users \(u\) and \(v\) is the overlap between
their liked-content sets:
\begin{equation}
    s_{\mathrm{UCF}}(u,v)
    =
    |L_u\cap L_v|.
    \label{eq:ucf-similarity}
\end{equation}
The \(50\) users with the largest overlap with \(u\) form the neighborhood
\(\mathcal{N}_{50}(u)\). A candidate content \(p\) is then scored as
\begin{equation}
    \operatorname{score}_{\mathrm{UCF}}(u,p)
    =
    \sum_{v\in\mathcal{N}_{50}(u)}
    \mathbf{1}\!\left[p\in L_v\right].
    \label{eq:ucf-score}
\end{equation}
The overlap magnitude selects the neighborhood but does not weight the
score: every neighbor contributes equally. Contents already liked by
\(u\) and contents authored by \(u\) are excluded from the candidate set.
Ties are broken by reverse-chronological order.
{\color{black} When \(u\) has not yet liked any content, the implementation
falls back to random selection.}

\paragraph{Linear Ranker.}
The Linear Ranker (LR) follows a two-stage recommendation procedure. In
the first stage, it constructs a candidate set from three pools:
\emph{(i)} a reverse-chronological follower-aware pool;
\emph{(ii)} a global popularity pool; and
\emph{(iii)} a UCF pool. The follower-aware pool uses a follower-content
quota of \(0.6\). For a requested timeline size \(k\), each pool contains
at most
\begin{equation}
    M=\min(10k,100)
    \label{eq:lr-pool-size}
\end{equation}
candidates{\color{black}, and the three pools are then merged and
deduplicated}.
{\color{black} Each pool applies its own top-up rule when it cannot fill
\(M\) slots: the popularity pool completes itself in reverse-chronological
order, the follower-aware pool draws from the global catalog, and the UCF
pool returns randomly selected contents when it is empty. The candidate set
is therefore not strictly the union of the three ranking criteria.}

In the second stage, each candidate content \(p\) is assigned the score
\begin{equation}
\begin{aligned}
S_{\mathrm{LR}}(u,p)={}&
0.28\,R(p)
+0.25\,F(u,a_p)
+0.15\,A(u,a_p)\\
&+0.08\,A_{\mathrm{recent}}(u,a_p)
+0.16\,T(u,p)\\
&+0.08\,\sigma(u,a_p),
\end{aligned}
\label{eq:lr-score}
\end{equation}
where \(u\) is the agent who queried the feed and \(a_p\) is the creator
of \(p\).

The recency component is
\begin{equation}
    R(p)
    =
    \exp\left(-\frac{\Delta_p}{10}\right),
    \label{eq:lr-recency}
\end{equation}
where \(\Delta_p\) is the age of \(p\), measured in simulation rounds.
The followed-author component is
\begin{equation}
F(u,a_p)
=
\begin{cases}
1, & \text{if \(u\) follows \(a_p\)},\\
0, & \text{otherwise.}
\end{cases}
\label{eq:lr-follow}
\end{equation}

Let \(n_{\mathrm{likes}}(u,a_p)\) denote the number of contents created
by \(a_p\) and previously liked by \(u\). In the implementation used
in our experiments, user--author affinity is
\begin{equation}
    A(u,a_p)
    =
    \log\left(
        1+n_{\mathrm{likes}}(u,a_p)
    \right).
    \label{eq:lr-affinity}
\end{equation}%
The recent-affinity component is a fixed discounted version of the same
quantity:
\begin{equation}
    A_{\mathrm{recent}}(u,a_p)
    =
    0.5\,A(u,a_p).
    \label{eq:lr-recent-affinity}
\end{equation}

\noindent \textit{{\upd Topic representation.}} Let \(\mathcal{T}\) denote the global topic taxonomy, with \(|\mathcal{T}|=10\), and let \(P_p\subseteq\mathcal{T}\) be the set of labels assigned to content \(p\). {\upd Each agent is initialized with a random subset of topics from \(\mathcal{T}\), and its topic profile subsequently evolves through publishing and commenting activity.} {\upd For content generation, the relevance of each topic is determined by the agent's publishing and commenting activity within the previous \(336\) rounds} (\(14\) simulated days){\upd : topics encountered more frequently in this window are more likely to be selected for subsequent posts, whereas topics with no recent activity progressively cease to influence content generation.}  {\upd New posts sample one topic proportionally to these recent interaction frequencies, while comments inherit the topic of the content to which they reply. Commenting on a previously unseen topic therefore introduces that topic into the agent's interaction history. Hence,} \(|P_p|=1\) for every content item. 
{\upd For the Linear Ranker, \(I_u\subseteq\mathcal{T}\) denotes the cumulative set of topics with which user \(u\) has interacted up to the current round. Unlike the recent topic frequencies used for content generation, \(I_u\) does not forget past topics and can therefore only expand over time.} The LR computes topic similarity as the Jaccard similarity between the target user's interest labels and the candidate content's topic labels{\color{black}, the only {\upd stage of the recommendation process} {\upd in} which topic
metadata {\upd are used}}: \begin{equation} T(u,p)=\frac{|I_u\cap P_p|}{|I_u\cup P_p|}. \label{eq:lr-topic-similarity} \end{equation}

Finally, two users are considered similar when they have liked at least one common content. Let \(c(u,a_p)\) denote the number of users who share at least one liked content with \(u\) and follow creator \(a_p\). The similar-user author signal is
\begin{equation}
    \sigma(u,a_p)
    =
    \log\left(
        1+c(u,a_p)
    \right).
    \label{eq:lr-similar-user}
\end{equation}
{\color{black}
The query considers at most \(100\) qualifying users, without ranking by overlap magnitude. 
{\upd When more than \(100\) users qualify, the selected users therefore form an arbitrary subset rather than the \(100\) users with the highest overlap.}}
Candidates are ranked by decreasing \(S_{\mathrm{LR}}(u,p)\), and the top-\(k\) contents are returned. {\color{black} Unlike UCF, the LR does not exclude contents already liked by
\(u\) from the candidate set. If the candidate set is empty, or if fewer than \(k\) candidates survive scoring, the feed is filled with randomly selected posts.}

The LR feature weights, pool-size thresholds, and discount factors are fixed
YSocial implementation choices and were not calibrated against empirical user
behavior. We therefore treat LR as one illustrative multi-signal
configuration rather than as a representative model of a deployed ranking
system. Its results are specific to the configuration described above and
should not be generalized to alternative feature weights.

{\upd Two aspects of the LR parametrization are relevant for interpreting its score. First, the feature terms operate on different scales, so the nominal coefficients should not be interpreted directly as measures of relative feature importance. Second, some components provide only limited independent variation in the configuration used here: because each content has a single topic, the topic-similarity contribution is small and decreases as the user's cumulative topic set expands, while \(A_{\mathrm{recent}}=0.5A\) is a fixed rescaling of the affinity term rather than an independent signal. We retain the default YSocial parametrization in our experiments and therefore interpret LR as one specific multi-signal ranking configuration rather than as a calibrated combination of equally weighted features.}

\FloatBarrier
\section{Degree-Resolved Visibility Under Global Recommenders}
\label{sec:appendix-global-degree}

\begin{figure}[ht]
\centering
\includegraphics[width=\linewidth]{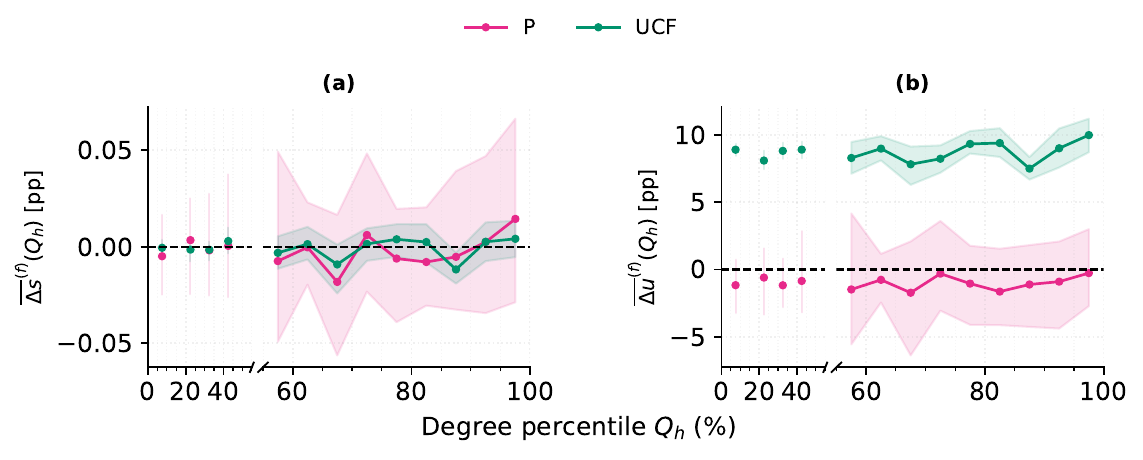}
\caption{%
  Changes in creator visibility by degree for global recommenders relative to the reverse-chronological baseline~RC {\vp in BA topology}.
  Panel~(a) reports changes in mean recommendation share {\vp ($\overline{\Delta s}^{(f)}(Q_h)$)}
  and panel~(b) reports changes in mean unique reach {\vp ($\overline{\Delta u}^{(f)}(Q_h)$)},
  both relative to RC and expressed in percentage points.
  {\upd Binning, axis break, and shaded areas follow Figure~\ref{fig:ba-degree-amplification}.}
}
\label{fig:appendix-global-degree-ba-er}
\end{figure}

Figure~\ref{fig:appendix-global-degree-ba-er} reports degree-resolved creator visibility for global recommenders in BA  topology, complementing the network-aware analysis in Figure~\ref{fig:ba-degree-amplification} of the main text.

\FloatBarrier

Global recommenders do not use the follower graph for candidate selection, so they do not create the structural coupling between creator degree and visibility that characterizes FP.
{\upd Under P, both recommendation share and unique reach stay close to the RC baseline across the whole degree-percentile range, with no monotonic gradient. UCF leaves recommendation share essentially unchanged but sits uniformly above RC in unique reach, by roughly 7--10~percentage points, again without a degree-linked trend.}
These results support the interpretation that the degree-linked amplification observed under FP is a product of the joint action of popularity ranking and the follower graph, not of popularity ranking alone.

\section{Pairwise Baseline Comparisons}
\label{sec:appendix-comparison}
\begin{figure}[ht]
\centering
\includegraphics[width=\linewidth]{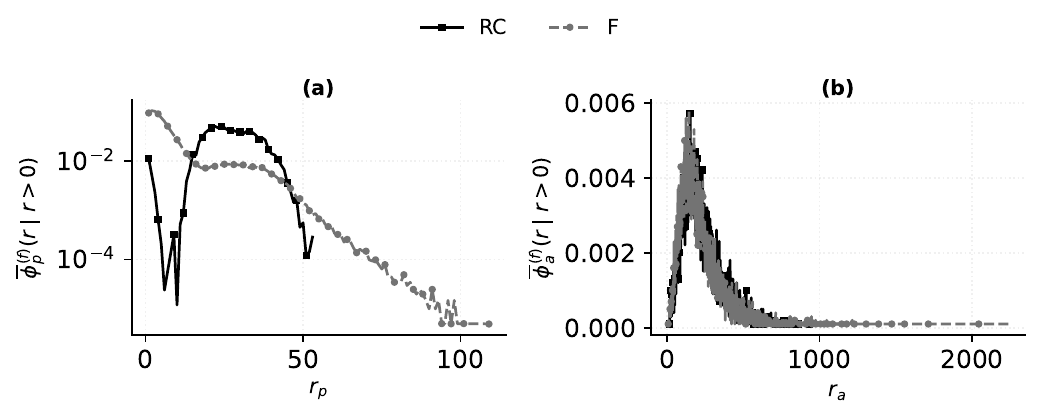}
\caption{%
  Visibility distributions for the reverse-chronological baseline~RC
  and the follower-only baseline~F.
  Panel~(a): content visibility distribution in the BA topology, {\vp \(\overline{\phi}_p^{(f)}(r\mid r>0)\)}.
  Panel~(b): creator visibility distribution in the BA topology, {\vp \(\overline{\phi}_a^{(f)}(r\mid r>0)\)}.
}
\label{fig:appendix-RC_vs_F}
\end{figure}

This appendix provides a pairwise comparison that motivates design choices discussed in the main text: the comparison between the two baselines RC and F (Figure~\ref{fig:appendix-RC_vs_F}), which
establishes that they are not equivalent and justifies treating them separately.

\paragraph{RC vs.\ F.}

\FloatBarrier

At the content level, the two baselines differ clearly.
Under RC, most of the probability mass is concentrated in a moderate-exposure
range around $r_{\mathrm{content}} \simeq 20$--$40$, with only a small residual
mass at very low visibility values.
Under F, the distribution decreases more smoothly and extends over a wider range, with higher mass in the low-to-intermediate region and a
longer right tail.
This reflects the effect of the follower constraint: restricting the
candidate pool to followed creators surfaces contents that would otherwise
age out under a global stream, so that far more contents receive at least
one recommendation, each at a lower exposure volume.
At the creator level, RC and F produce largely overlapping distributions.
Both concentrate most probability mass in a similar visibility range,
and the curves differ only marginally around the peak; they separate only
in the upper tail, which is heavier under F.
This convergence occurs because creator visibility aggregates over
multiple contents, averaging out the content-level differences between
the two baselines.
Neither baseline amplifies creators through popularity or collaborative
signals, which further limits their divergence at the creator level.
A further property of RC is that visibility loss is driven by slot
competition rather than poor timing: peak-activity hours increase the
probability that a post receives no impressions, while contents that do
break through at those hours receive more exposure, so that the two
effects compensate and the expected exposure of a post stays nearly flat
across posting hours.
Creators who post more frequently therefore accumulate more
recommendation chances regardless of when they post.

{\vm
\section{Topology Ablation}
\label{sec:appendix-network-ablation}
{\vp

\subsection{Networks Description.}
{\color{black}
{\upd We consider three generative network models to represent different network topologies. To ensure comparability,}
at each population size, {\upd their} generative parameters are
{\upd selected} so that the resulting networks have approximately the same
{\upd density across all population sizes. For a given \(N\), this also yields comparable mean degree across topologies, while mean degree increases as \(N\) increases.}}

\noindent \textbf{Erd\H{o}s--R\'enyi random network (ER).} The Erd\H{o}s--R\'enyi (ER) network is generated according to the \(G(N,p_{\mathrm{ER}})\) model \citep{erdhos1959random}, in which each unordered pair of nodes is connected independently with probability \(p_{\mathrm{ER}}=0.02\). Consequently, node degrees follow a binomial distribution, with expected degree 
 \(\mathbb{E}[k]=(N-1)p_{\mathrm{ER}}\). 
{\color{black} The same connection probability is used at every population size,
so density is constant by construction and mean degree grows linearly with \(N\).}
Compared with the reference BA model, the resulting degree distribution is relatively narrow around its mean and does not generate the pronounced hubs.  

{\color{black} \noindent \textbf{Barab\'asi--Albert network (BA).} The reference
BA network is generated with attachment parameter \(m\), which fixes the edge
count at \(m(N-m)\). At each population size, \(m\) is set to the value that
matches the density of the corresponding ER network, giving \(m=10\), \(50\) and
\(100\) for \(N=1{,}000\), \(5{,}000\) and \(10{,}000\).}

\noindent \textbf{Power-law cluster network (PC).} The power-law cluster (PC) network is generated using the Holme--Kim model \cite{holme2002growing}. The model extends the Barab\'asi--Albert preferential-attachment mechanism by introducing triadic closure, modulated by a triangle-formation probability \(p_{\mathrm{tri}}\). 
The resulting network therefore preserves the heterogeneous, approximately power-law degree structure induced by preferential attachment while allowing substantially greater local clustering than the standard Barab\'asi--Albert model. {\color{black} PC networks use the same \(m\) as the corresponding BA network and a triangle-formation probability \(p_{\mathrm{tri}}=0.5\), which is held constant across population sizes.}

A summary of generated network statistics can be found in Table \ref{tab:network_metrics}. 

\begin{table}[hbtp]
\centering
\caption{Structural properties of the generated networks{\color{black}, together
with the generative parameters used at each population size. 
Values refer to the network instances used in the simulations.}}
\label{tab:network_metrics}
\resizebox{\linewidth}{!}{%
\begin{tabular}{llcrrrrr}
\toprule
Topology &
$N$ &
{\color{black} Parameters} &
$|E|$ &
$k_{\min}$ &
$k_{\max}$ &
$\langle k \rangle$ &
Density \\
\midrule

ER & 1,000  & {\color{black} $p_{\mathrm{ER}}=0.02$} & 10,127  & 7   & 37    & 20.25  & 0.02027 \\
ER & 5,000  & {\color{black} $p_{\mathrm{ER}}=0.02$} & 248,978 & 64  & 136   & 99.59  & 0.01992 \\
ER & 10,000 & {\color{black} $p_{\mathrm{ER}}=0.02$} & 999,269 & 150 & 253   & 199.85 & 0.01999 \\

\midrule

BA & 1,000  & {\color{black} $m=10$}  & 9,900   & 10  & 182   & 19.80  & 0.01982 \\
BA & 5,000  & {\color{black} $m=50$}  & 247,500 & 50  & 804   & 99.00  & 0.01980 \\
BA & 10,000 & {\color{black} $m=100$} & 990,000 & 100 & 1,623 & 198.00 & 0.01980 \\

\midrule

PC & 1,000  & {\color{black} $m=10$, $p_{\mathrm{tri}}=0.5$}  & 9,831   & 8  & 235   & 19.66  & 0.01968 \\
PC & 5,000  & {\color{black} $m=50$, $p_{\mathrm{tri}}=0.5$}  & 245,836 & 48 & 953   & 98.33  & 0.01967 \\
PC & 10,000 & {\color{black} $m=100$, $p_{\mathrm{tri}}=0.5$} & 980,630 & 97 & 3,197 & 196.13 & 0.01961 \\

\bottomrule
\end{tabular}%
}
\end{table}

\subsection{Results}

We compare the BA reference network with the ER and PC topologies to assess their role in determining how the visibility patterns are shaped. 
The comparison reported in this section is restricted to the reference simulation setting, i.e., \(D=30\) days, \(N=1000\) agents, feed size \(k=12\), and visibility window \(\tau_{\ell}=36\) rounds, with all remaining parameters fixed as in the main experiments.
Results are based on $J=10$ runs per topology and recommender. 
Topology effects {\upd while varying duration, population size, visibility window or feed size} are discussed in the OFAT analysis (Appendix \ref{app:ofat}). {\upd Further} results for the full set of sensitivity configurations are available for inspection in the online dashboard\footnote{\url{https://ofat-sensitivity-dashboard.github.io/}}. 

The results show an intuitive distinction between global and network-aware recommenders: {\upd outcomes under} global recommenders are insensitive to topology changes, whereas {\upd with} network-aware recommenders {\upd outcomes} exhibit small but systematic effects.


\smallskip

\noindent Topology has a small, mostly quantitative, effect, in the reference setting.

\noindent \textbf{Content visibility.} At the content level,  {\upd topology produces only modest changes in the recommendation-volume distributions (\(\phi_p\)) of the network-aware recommenders. ER tends to show a more compressed upper tail than BA and PC, particularly for FP and LR, while BA and PC remain closely aligned} (Figure~\ref{fig:network-ablation-phi-p}). 

\begin{figure}[t]
\centering
\includegraphics[width=0.9\linewidth]{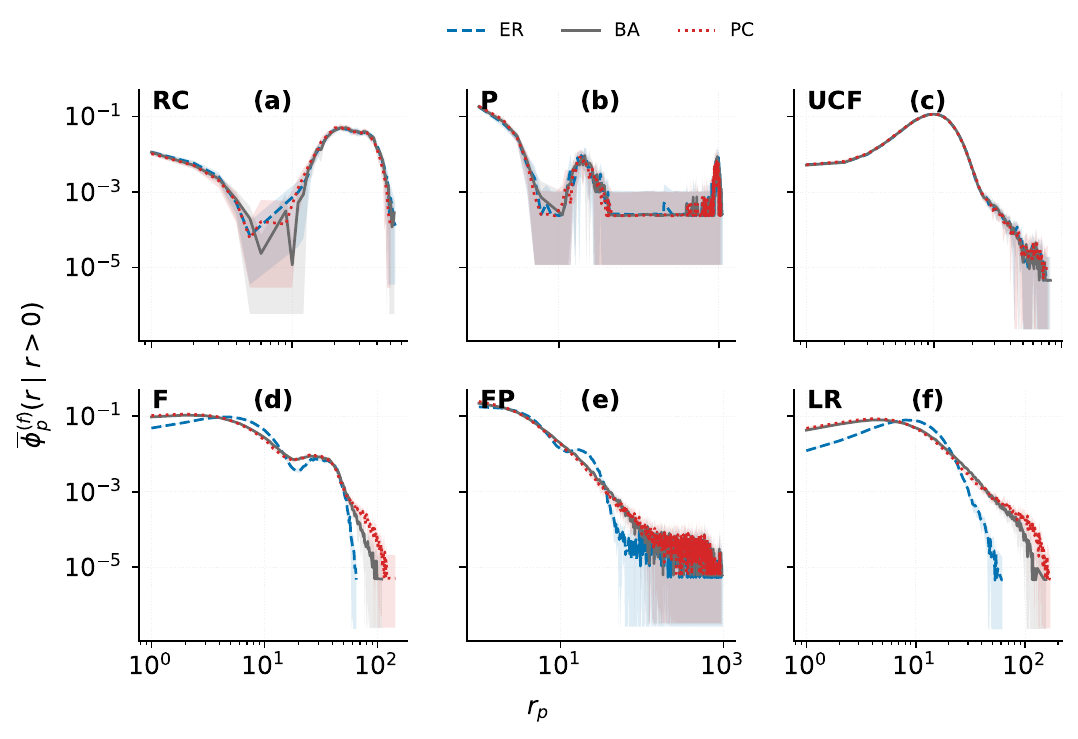}
\caption{\textbf{Content-level recommendation-volume distributions across network topologies.} Each panel reports the mean of \(\phi_p^{(f)}(r\mid r>0)\), with shaded regions indicating standard deviation. 
The top row reports global recommenders (RC, P, and UCF) and the bottom row network-aware recommenders (F, FP, and LR). 
Curves compare {\upd ER (blue), BA (baseline, grey), and PC (red)} networks under the reference simulation setting.}
\label{fig:network-ablation-phi-p}
\end{figure}

The aggregate metrics make differences {\upd between topologies} more apparent (Figure~\ref{fig:network-ablation-scalars}). 
For the network-aware recommenders F, FP and LR, the degree-heterogeneous
BA and PC topologies generally produce greater content-level inequality than ER,
indicating that heterogeneity in network position translates into a more uneven
allocation of recommendation volume. This can also reduce content coverage, most
clearly under FP, where popularity ranking operates within follower-based
candidate pools.
LR illustrates that these two effects need not coincide: its recommendation-volume distribution and \(G_p\) change substantially between ER and BA/PC, while \(C_p\) remains close to complete coverage, meaning that topology redistributes exposure without excluding contents from recommendation altogether. 
\begin{figure}[hbtp!]
\centering
\includegraphics[width=0.8\linewidth]{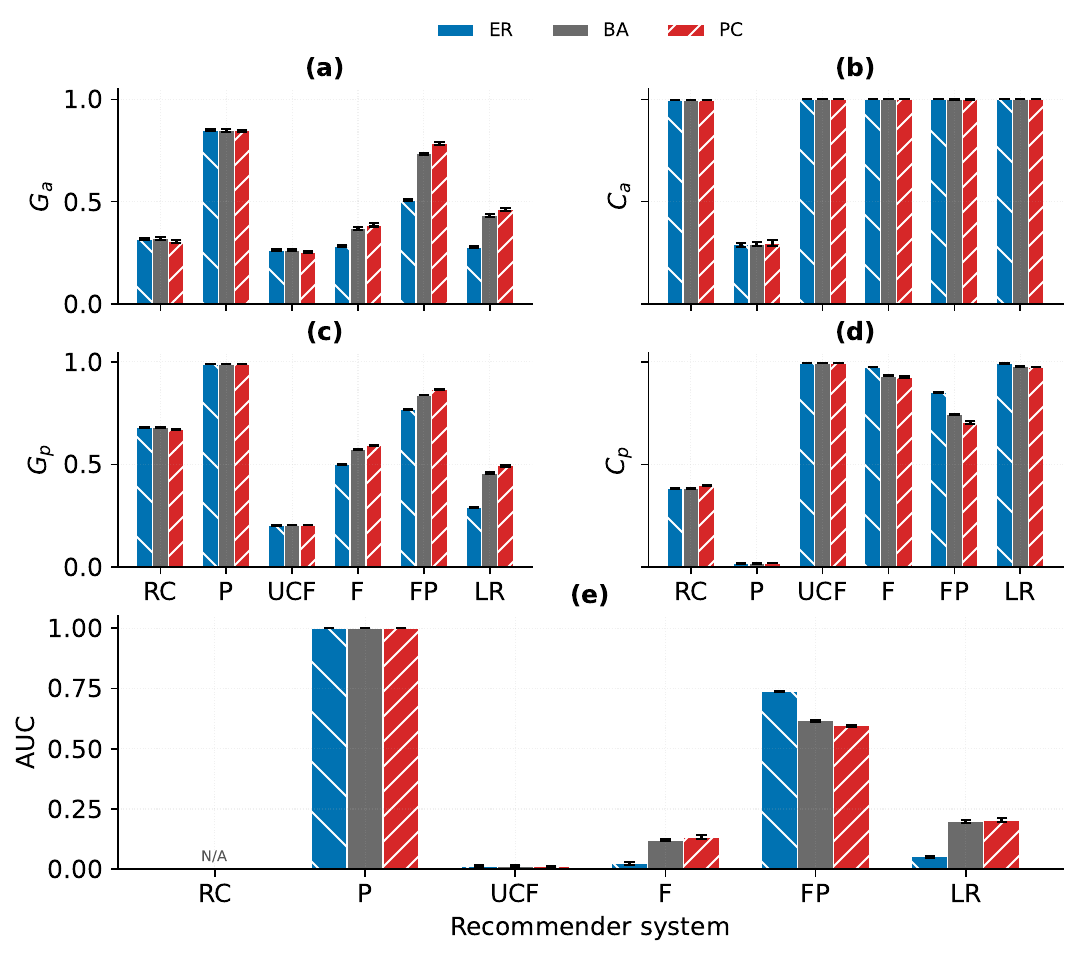}
\caption{\textbf{Absolute concentration, coverage, and popularity reinforcement across network topologies.} Bars report means of the absolute  Gini {\upd for creators (a) and content (c)}, coverage {\upd for creators (b) and content (d)}, and popularity reinforcement AUC {\upd (e)}, with error bars indicating sample standard deviation. Topologies are ordered ER (blue), BA (grey), and PC (red). 
Popularity-reinforcement AUC is not reported for RC, whose correlation is defined over at most the first two content ages.}
\label{fig:network-ablation-scalars}
\end{figure}
{\upd Temporal reinforcement does not follow the same topology dependence as static content inequality. 
Under FP, popularity reinforcement is stronger in ER than in BA and PC, despite ER producing lower content inequality and higher coverage. 
A possible interpretation is that the narrower degree distribution of ER reduces differences in exposure opportunities associated with creator network position, making accumulated reactions relatively more important for determining subsequent exposure. 
In BA and PC, degree heterogeneity introduces an additional source of visibility differences, weakening the relative contribution of popularity to future exposure (see Creator visibility). 
F and LR show the opposite AUC trend, indicating that the interaction between topology and temporal reinforcement depends on the recommendation mechanism rather than following a universal effect of degree heterogeneity.}

\begin{figure}[h]
\centering
\includegraphics[width=0.9\linewidth]{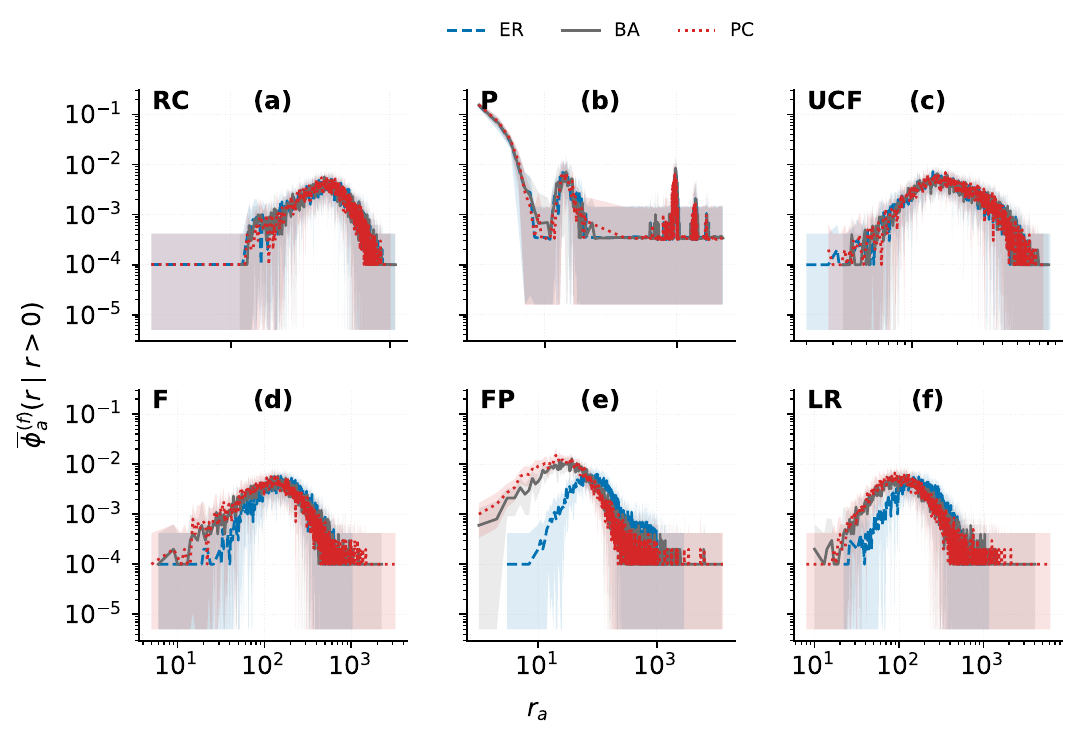}
\caption{\textbf{Creator-level recommendation-volume distributions across network topologies.} Each panel reports the mean of \(\phi_a^{(f)}(r\mid r>0)\), with shaded regions indicating standard deviation.  The top row reports global recommenders (RC, P, and UCF) and the bottom row network-aware recommenders (F, FP, and LR). Curves compare ER, BA, and PC networks under the reference simulation setting.}
\label{fig:network-ablation-phi-a}
\end{figure}

\smallskip
\noindent \textbf{Creator visibility.}
At the creator level, the recommendation-volume distributions {\upd ($\phi_a$)} show a slightly higher prevalence of low-exposure creators under BA and PC than under ER (Figure \ref{fig:network-ablation-phi-a}). 
The BA and PC curves largely overlap, indicating that the additional triadic closure introduced by PC does not substantially change the overall shape of the visibility distribution.

Despite this similarity, creator-level concentration ($G_a$) is slightly higher under PC than under BA, while both {\upd concentrate visibility} more than ER. 
Creator coverage, however, remains essentially complete across all three topologies, indicating that topology {\upd changes how total exposure is distributed across creators without substantially affecting whether creators receive any exposure at all.}

{\upd The relationship between recommendation share \(\overline{s}^{(f)}(Q_h)\), or unique reach \(\overline{u}^{(f)}(Q_h)\), and creators' degree offers the} clearest view of how {\upd visibility is redistributed among creators under different network topologies (Figures~\ref{fig:network-ablation-degree-s}(d)--(f) and~\ref{fig:network-ablation-degree-u}(d)--(f))}.

\begin{figure}[hbtp!]
\centering
\includegraphics[width=0.9\linewidth]{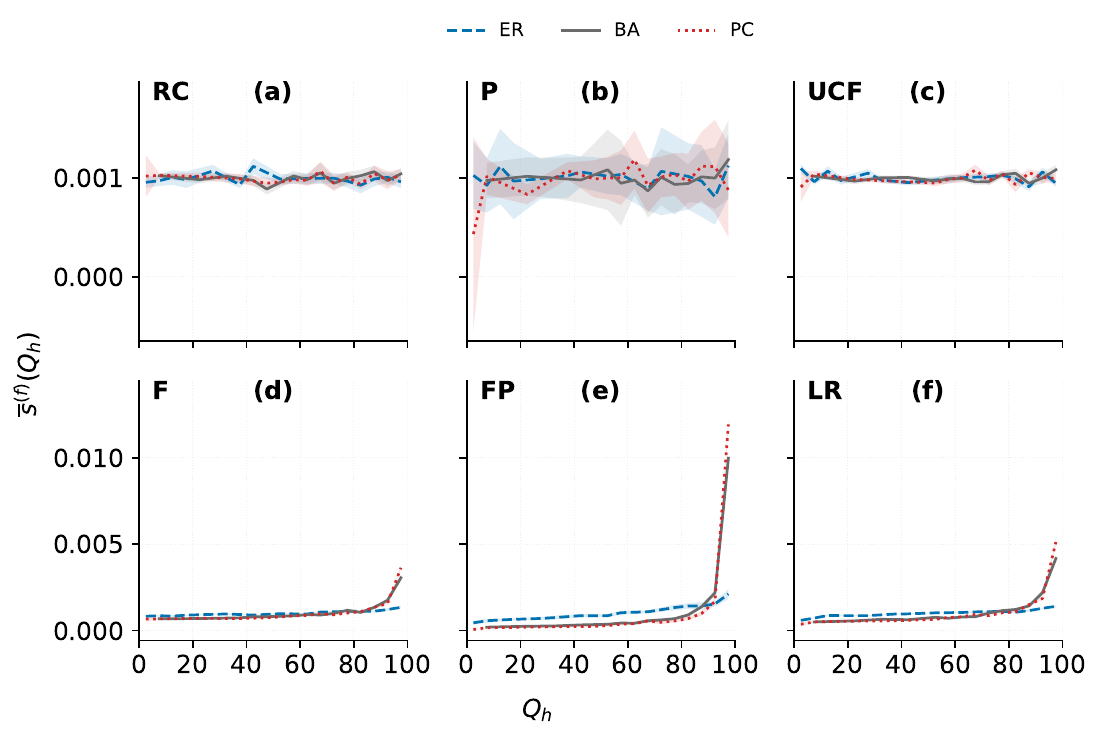}
\caption{\textbf{{\upd Recommendation share by degree.}} Mean creator recommendation share \(\overline{s}^{(f)}(Q_h)\) is reported across 20 tie-preserving creator-degree percentile bins. Curves compare {\upd ER (blue), BA (baseline, grey), and PC (red)} networks under the reference simulation setting.}
\label{fig:network-ablation-degree-s}
\end{figure}
}
BA and PC produce nearly overlapping curves in F, FP and LR,  {\upd whereas ER redistributes visibility more evenly across degree classes. 
Under F, the three topologies remain close through most of the degree distribution, with a clear separation emerging only in the highest-degree ventiles (approximately \(Q_h\gtrsim80\)): in the top ventile, BA and PC reach roughly twice the recommendation share and about \(1.5\)--\(2\times\) the unique reach observed under ER. 
The contrast is strongest under FP. 
\begin{figure}[hbtp!]
\centering
\includegraphics[width=0.9\linewidth]{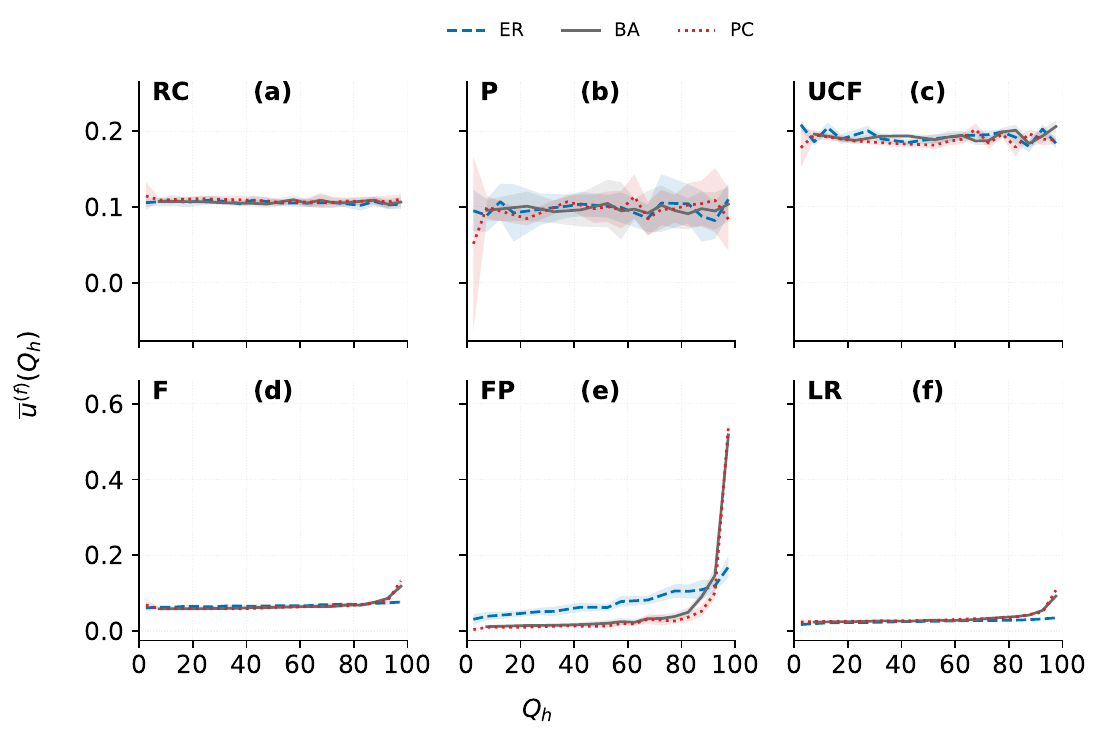}
\caption{\textbf{{\upd Unique reach by degree.}} Mean normalized creator reach \(\overline{u}^{(f)}(Q_h)\) is reported across 20 tie-preserving creator-degree percentile bins.  Curves compare {\upd ER (blue), BA (baseline, grey), and PC (red)} networks under the reference simulation setting.}
\label{fig:network-ablation-degree-u}
\end{figure}
Across low- and intermediate-degree ventiles, ER creators receive equal or greater visibility than under the BA reference, but the curves reverse sharply at the upper end of the degree distribution. 
In the highest ventile, recommendation share under BA and PC is approximately \(5\times\) that under ER, while unique reach is roughly \(3\times\) larger. 
LR shows the same redistribution but less strongly: ER is comparable to or above BA over much of the lower and middle degree range, whereas BA and PC increasingly dominate above approximately the 80th--85th percentile; in the final ventile, their recommendation share is around \(3\)--\(4\times\) and unique reach around \(2\)--\(3\times\) the ER values. 
Thus, the weaker degree gradient under ER reflects both greater visibility for lower-degree creators and substantially less amplification of the highest-degree creators relative to the BA reference. 
This is consistent with ER's narrower degree distribution providing less structural heterogeneity for follower-based candidate selection to amplify. 
The near overlap of BA and PC across ventiles, including the sharp increase at the upper tail, further suggests that the additional clustering introduced by PC has comparatively little effect on degree-resolved visibility once a similarly heterogeneous degree structure is present.}}

\begin{figure}[hbtp!]
\centering
\includegraphics[width=\linewidth]{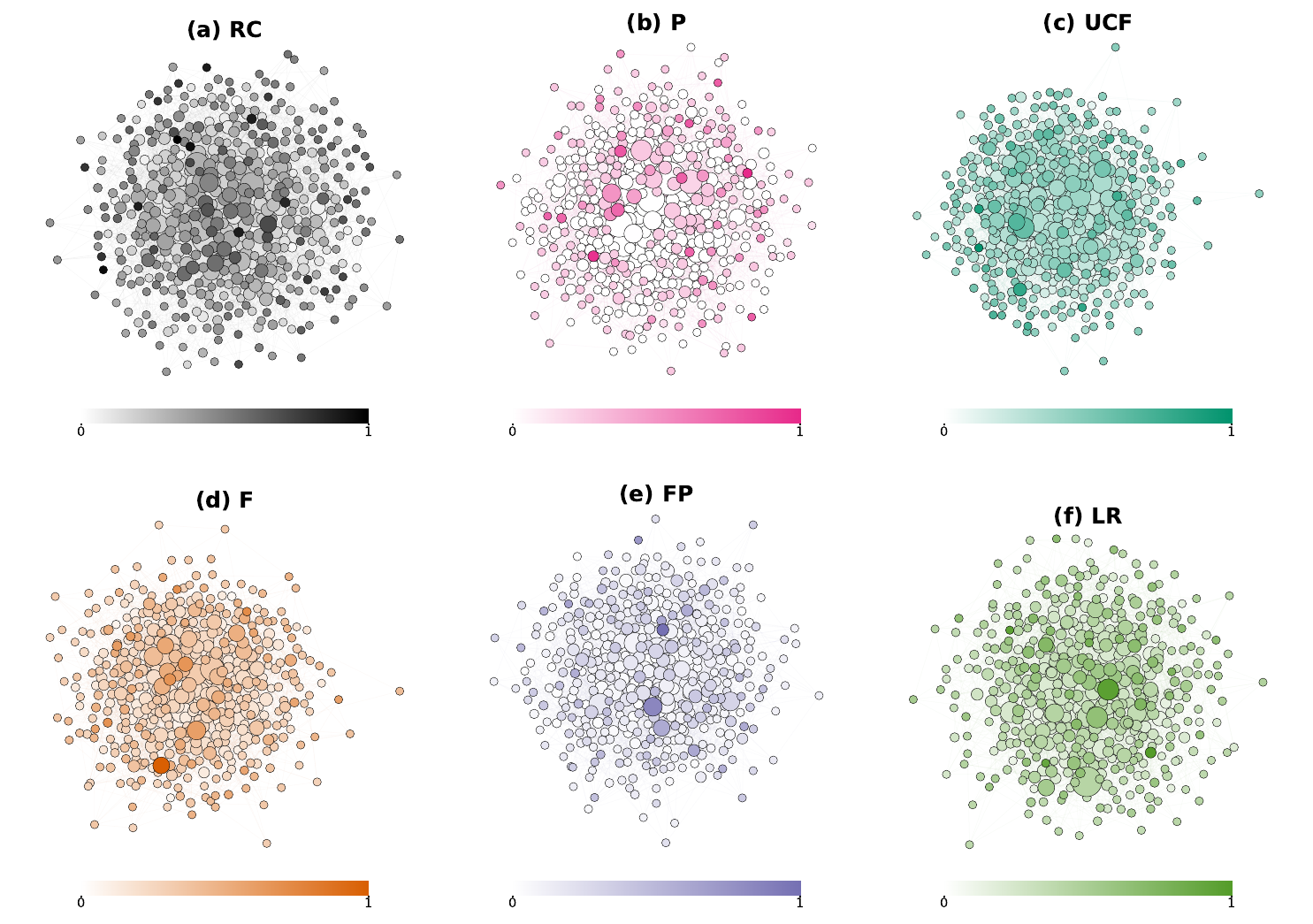}
\caption{\textbf{Creator visibility for each recommender system (Erd\H{o}s--R\'enyi network).} Each panel shows creator visibility in the ER social network under a different feed: (a)~RC, (b)~P, (c)~UCF, (d)~F, (e)~FP, (f)~LR. Node size is proportional to degree; node saturation reflects normalized visibility. Each panel shows the union of the 300 nodes with the highest degree and the 300 creators with the highest recommendation exposure.}
\label{fig:network-ablation-topology-er}
\end{figure}

\begin{figure}[hbtp]
\centering
\includegraphics[width=\linewidth]{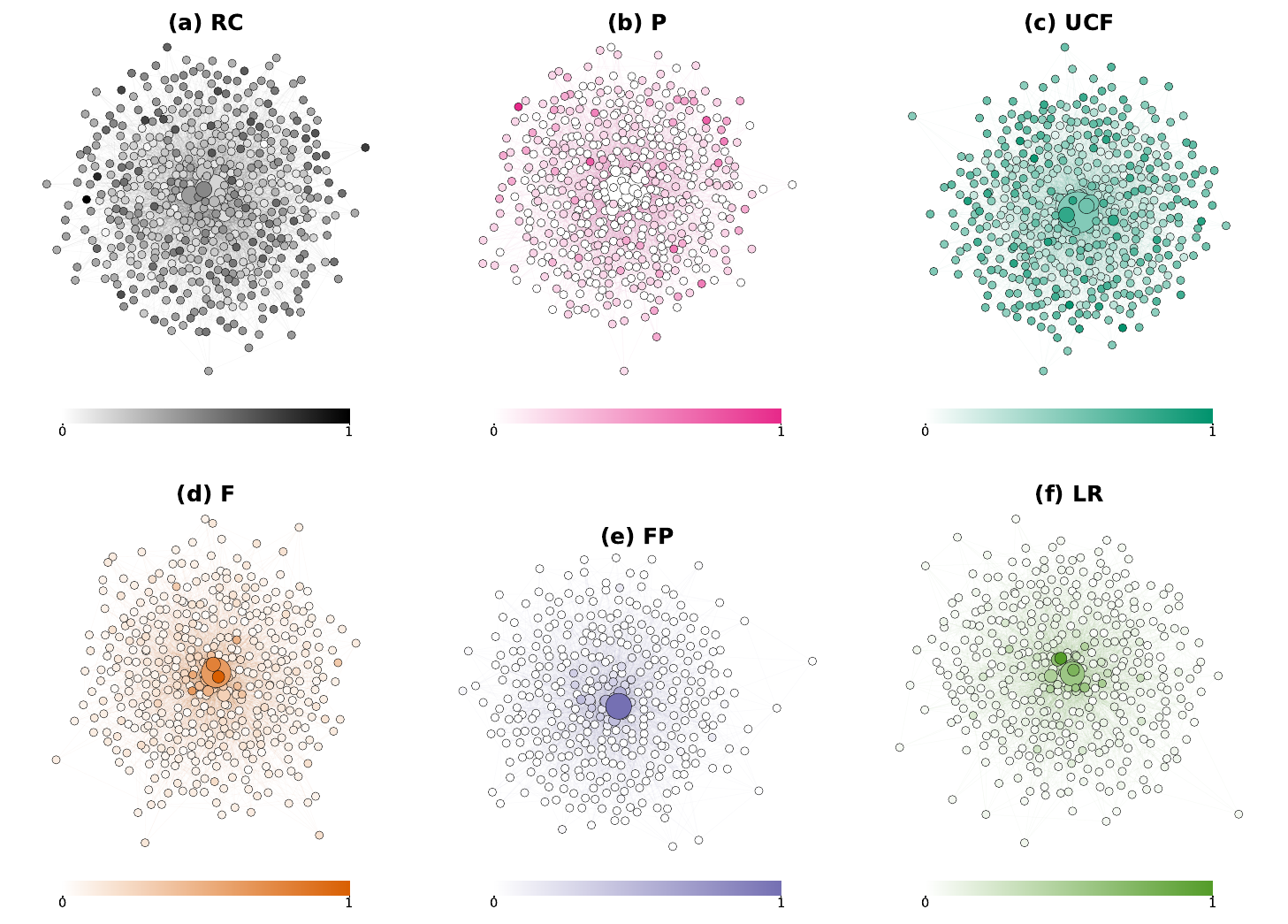}
\caption{\textbf{Creator visibility for each recommender system (power-law-cluster network).} Each panel shows creator visibility in the PC social network under a different feed: (a)~RC, (b)~P, (c)~UCF, (d)~F, (e)~FP, (f)~LR. Node size is proportional to degree; node saturation reflects normalized visibility. Each panel shows the union of the 300 nodes with the highest degree and the 300 creators with the highest recommendation exposure.}
\label{fig:network-ablation-topology-pc}
\end{figure}

{\color{black} The same pattern is visible directly on the graphs
(Figures~\ref{fig:network-ablation-topology-er}
and~\ref{fig:network-ablation-topology-pc}), which reproduce
Figure~\ref{fig:ba-network-viz} for the ER and PC topologies. Under ER,
follower-popularity spreads visibility over a broad set of nodes, whereas
under PC it concentrates it on a few hubs, as in BA.}

\section{OFAT Analysis}\label{app:ofat}
{\upd This section reports the detailed results of the OFAT sensitivity analysis, organized by the parameter varied. For each parameter, we summarize Gini concentration, coverage, and popularity-reinforcement AUC in Tables~\ref{tab:sensitivity-summary-visibility}, ~\ref{tab:sensitivity-summary-duration},  ~\ref{tab:sensitivity-summary-feed}, and~\ref{tab:sensitivity-summary-population}. 
As in Figure~\ref{fig:network-ablation-scalars}, popularity-reinforcement AUC is not reported for RC.
Additional results, including the complete set of sensitivity configurations and complementary visualizations, are available in the online dashboard\footnote{\url{https://ofat-sensitivity-dashboard.github.io/}}.}

\subsubsection{Sensitivity to Visibility Window ($\tau_\ell$).}

\paragraph{Content visibility.} 
Sensitivity to the visibility window is localized to popularity-based feeds (P and FP), primarily affecting coverage and popularity reinforcement.
As we can see from the aggregate metrics in Table~\ref{tab:sensitivity-summary-visibility}, as $\tau_\ell$ increases, content coverage decreases, most visibly for FP.
Concentration {\upd ($G_p$)} is essentially unchanged, with only very small negative changes for $\tau_\ell=24$ and very small positive changes for $\tau_\ell=48$. 
{\upd Under} P {\upd concentration} moves in the same direction on a compressed scale ({\upd since it is} already near its  {\upd maximum value} at every setting).
RC, UCF, F, and LR vary only marginally.
The pattern is consistent across topologies.

Popularity reinforcement follows the same pattern, {\upd decreasing} most visibly for FP at the shortest window, consistent with less time for an early-popularity advantage to compound. Lengthening the window beyond the reference adds little, so the effect is concentrated between \(\tau_\ell=24\) and the reference setting.

These changes are consistent with a persistence--turnover interpretation: a longer visibility window allows already-successful content to remain eligible for recommendation for longer, whereas a shorter window forces more frequent replacement. This interpretation is also consistent with the stability of $\phi_p$ across settings: changing $\tau_\ell$ appears to affect primarily how many contents enter the exposed set, while leaving the distribution of recommendation volume among exposed contents largely unchanged.

\paragraph{Creator visibility.}
Sensitivity at the creator level is also concentrated in popularity-based feeds, and is strongest under P.
As we can see from the aggregate metrics in Table~\ref{tab:sensitivity-summary-visibility}, as $\tau_\ell$ increases, creator visibility becomes more concentrated and creator coverage decreases under P, while FP follows the same direction more moderately in concentration while retaining near-complete creator coverage. 
RC, UCF, F, and LR vary only marginally, with the exception of a small shift in LR concentration between \(\tau_\ell=24\) and the reference.
The degree-resolved patterns are instead essentially unchanged across visibility-window settings.

These results are consistent with the same interpretation observed at the content level: longer-lived successful content can continue directing exposure toward the same creators, increasing aggregate creator-level concentration, while shorter visibility windows increase turnover and create more opportunities for different creators to receive exposure. 
The stability of the degree-resolved curves further suggests that $\tau_\ell$ affects the overall {\upd concentrating behavior of the recommender system,} rather than how visibility is distributed across network positions.


{\vp
\subsubsection{Sensitivity to Simulation Duration $D$}
\label{sec:sensitivity-duration}


\paragraph{Content visibility.}
Content-level visibility is essentially invariant across simulation durations, for all recommenders and topologies.
As shown by the aggregate metrics in Table~\ref{tab:sensitivity-summary-duration}, changes in content concentration and coverage remain negligible across $D\in\{30,60,90\}$ (in BA, $G_p$ changes by at most $0.03$, with $\Delta C_p = -7.67$ pp. for LR). Recommendation-volume distributions and popularity--future-exposure reinforcement are similarly stable, with P remaining near maximal reinforcement, FP retaining its intermediate reinforcement level, and UCF remaining close to zero. 
This stability is consistent with the fixed content-visibility window: increasing $D$ adds more complete content lifecycles to the simulation, but does not change how long an individual content remains eligible for recommendation or the exposure dynamics operating within that lifecycle.

\paragraph{Creator visibility.}
Creator-level visibility is more sensitive to simulation duration, with the clearest changes occurring under P.
As $D$ increases, creator concentration decreases and creator coverage increases substantially under P ($C_a$: $29.2\to50.3\to66.5$\% in BA), while RC follows the same direction more weakly. UCF, F, FP, and LR vary only marginally, as creator coverage is already close to complete under these recommenders. The same qualitative pattern is observed across topologies.
Because concentration under P decreases much more sharply than under FP as the simulation horizon increases, {\upd the main conclusion that follower-based filtering redistributes visibility relative to global popularity ranking no longer holds consistently across durations:} FP is the less concentrated of the two at the reference duration but the more concentrated one at $D=90$ in BA 
and already at $D=60$ in PC.
In ER, where FP is markedly less concentrated than P at the reference setting,
no reversal occurs within the tested range. 
The reversal concerns concentration only: creator coverage under P remains far below FP's at every duration. It is therefore driven by the horizon dependence of P rather than by a change in FP, and it leaves the content-level ordering unaffected.

These results are consistent with a cumulative-horizon interpretation: although P exposes only a restricted set of creators at any given time, extending the simulation allows a larger number of distinct creators to enter that exposed set over successive content lifecycles. 
By contrast, recommenders that already provide broad creator coverage have little additional visibility to gain from a longer observation horizon.
}


{\vp
\subsubsection{Sensitivity to Feed Size $k$}
\label{sec:sensitivity-feed-size}


\paragraph{Content visibility.}
Content-level visibility is broadly sensitive to feed size, especially for RC, F, and FP.

As shown by the aggregate metrics in Table~\ref{tab:sensitivity-summary-feed}, increasing $k$ decreases content concentration and increases coverage across recommenders and topologies. 
{\upd The magnitude of this effect is recommender-dependent: RC, F, and FP are the most sensitive, while P and UCF retain the most concentrated and broadly distributing regimes, respectively, and differences among intermediate strategies narrow as \(k\) grows.}
Popularity reinforcement {\upd remains essentially unchanged (LR decreases at \(k=24\): \(\Delta\mathrm{AUC}_\rho=-0.05\) in BA and \(-0.06\) in PC, relative to a baseline of 0.20), except for FP where it decreases with longer feeds.}

These results are consistent with an exposure-capacity interpretation: increasing \(k\) makes more recommendation slots available at each request, reducing competition for exposure and allowing a larger set of contents to enter the feed. This effect is strongest for recommenders whose output is constrained by a relatively narrow candidate set. {\upd Under FP, increasing \(k\) also exposes contents further down the popularity ordering, reducing the exclusivity of the most popular contents and weakening the association between early popularity and future exposure. Under global P, by contrast, popularity remains the ranking criterion itself, so a larger feed can broaden coverage without substantially weakening popularity reinforcement.}

\paragraph{Creator visibility.}
Creator-level visibility follows the same general response to feed size, although the magnitude of the effect differs across recommenders.
As $k$ increases, creator concentration decreases and creator coverage increases. This effect is clearer under P, while FP also becomes less concentrated but maintains near-complete creator coverage throughout. UCF, F, and LR vary only marginally because creator coverage is already close to saturation under these recommenders. The degree-resolved visibility patterns remain essentially unchanged across feed sizes, with FP and LR preserving the same degree-dependent gradients.

These results are consistent with feed size primarily changing the overall availability of exposure rather than its structural allocation across the network. Larger feeds create more opportunities for contents and creators to receive recommendations, but do not substantially alter the way visibility varies with creator degree.
}


\subsubsection{Sensitivity to Population Size}
\label{sec:sensitivity-population}
\paragraph{Content visibility.} Content-level visibility becomes increasingly
selective with system scale, although the mechanisms behind this trend differ
across recommenders. As shown by the aggregate metrics in
Table~\ref{tab:sensitivity-summary-population}, increasing \(N\) generally raises
content concentration and reduces coverage under RC, P, F, and FP. Because feed
size, visibility window, simulation duration, and network density remain fixed
while the amount of generated content grows approximately with \(N\), part of
this decline reflects the shrinking ratio between exposure capacity and catalog
size. This is clearest for RC, whose coverage closely follows inverse-\(N\)
scaling (\(C_p\): \(38.5\to7.8\to3.9\%\) in BA for
\(N=1{,}000\to5{,}000\to10{,}000\)), and which we therefore use as a reference
for interpreting the other recommenders. P is already near maximal concentration
at every scale, so its scale dependence appears only as further reductions in an
already-low coverage. More informative is FP, which moves toward the extreme
concentration of P while retaining substantially greater content coverage; F
responds in the same direction but less strongly, whereas UCF and LR remain
comparatively stable. Because density is held approximately constant, a larger
\(N\) also means a higher mean degree and larger follower neighborhoods, so this
experiment measures sensitivity to system scale rather than to population size
alone: both the catalog and, for network-aware recommenders, the candidate pools
grow relative to a fixed number of feed slots. Popularity reinforcement under FP
responds to the same pressure, increasing sharply with \(N\) and approaching the
near-maximal level of global P at the largest population tested, consistently
across topologies (\(\mathrm{AUC}_\rho\): \(0.62\to0.93\to0.97\) in BA,
\(0.74\to0.97\to0.99\) in ER, and \(0.60\to0.93\to0.96\) in PC), while P remains
near maximal and UCF near zero throughout. The attenuation of popularity
reinforcement observed under follower-based candidate selection at the reference
scale therefore does not persist as follower neighborhoods grow. These results
are consistent with a competition-for-exposure interpretation: popularity ranking
strengthens the effect by repeatedly promoting already-successful items, whereas
user-dependent recommenders such as UCF and LR can expose different subsets of
the enlarged catalog to different users.

\paragraph{Creator visibility.} Creator-level visibility shows {\upd a similar increase} in selectivity with scale, with the most evident change under popularity-based ranking. As \(N\) increases, creator concentration rises and creator coverage falls under P and FP. The change under FP is particularly important because near-complete creator coverage at the reference scale does not persist: in BA, coverage declines from approximately complete at \(N=1{,}000\) to \(88.4\%\) at \(N=10{,}000\), while creator concentration increases substantially; the decline is steeper in PC (\(C_a\) down to \(72.9\%\)) and milder in ER (\(95.7\%\)). Thus, follower-aware candidate selection does not by itself {\upd preserve} broad creator visibility at larger system scales. UCF {\upd and} LR instead retain near-complete creator coverage and comparatively stable creator-level concentration; F retains full creator coverage but becomes more concentrated. The same qualitative pattern is observed across topologies. {\upd The creator-level results are consistent with the same scaling mechanism observed for content. Under FP, popularity selects among increasingly large follower-based candidate pools while the number of feed slots remains fixed, allowing exposure to concentrate progressively on a smaller subset of creators. The increasing concentration observed under F, despite its complete creator coverage, further indicates that expanding follower neighborhoods affects the distribution of exposure independently of popularity.} {\upd Together, the content- and creator-level results show that sensitivity to \emph{system scale} is strongly recommender-dependent: globally shared rankings exhibit the expected pressure from a growing catalog, follower-based strategies additionally respond to expanding network neighborhoods, while UCF and LR remain comparatively stable across the tested population sizes.}

\begin{table*}[h!]
\centering
\caption{Sensitivity of Gini concentration, recommendation coverage, and popularity-reinforcement AUC to visibility window $\tau_\ell$. Non-reference columns show $\Delta$ vs.\ the block's reference setting; the reference column shows the absolute value in \textbf{bold}. $G_p$/$G_a$: content-/creator-level Gini, unitless $[0,1]$; $C_p$/$C_a$: content-/creator-level coverage (\%); AUC$_\rho$: interval-normalized area under the popularity$\to$future-exposure Spearman curve. AUC$_\rho$ is taken over each setting's own visibility window, so it is not comparable between settings; AUC$_\rho^{24}$ restricts every setting to the first 24 hours, the shortest window tested, and is the column to read when comparing across $\tau_\ell$.}
\label{tab:sensitivity-summary-visibility}
\resizebox{\textwidth}{!}{%
\small
\renewcommand{\arraystretch}{0.9}
\setlength{\tabcolsep}{4pt}
\begin{tabular}{ll r r r r r r r r r r r r r r r r r r}
\toprule
 &  & \multicolumn{6}{c}{$\tau_\ell=24$ } & \multicolumn{6}{c}{$\tau_\ell=36$ \textit{(ref.)}} & \multicolumn{6}{c}{$\tau_\ell=48$ } \\
\cmidrule(lr){3-8}\cmidrule(lr){9-14}\cmidrule(lr){15-20}
Topology & Recommender & $\Delta G_p$ & $\Delta G_a$ & $\Delta C_p$ (pp) & $\Delta C_a$ (pp) & $\Delta$AUC$_\rho$ & $\Delta$AUC$_\rho^{24}$ & $G_p$ & $G_a$ & $C_p$ (\%) & $C_a$ (\%) & AUC$_\rho$ & AUC$_\rho^{24}$ & $\Delta G_p$ & $\Delta G_a$ & $\Delta C_p$ (pp) & $\Delta C_a$ (pp) & $\Delta$AUC$_\rho$ & $\Delta$AUC$_\rho^{24}$ \\
\midrule
\multicolumn{20}{l}{\textit{BA topology}} \\
\midrule
 & RC & $-$0.01 & $-$0.01 & $+$1.00 & $+$0.04 & --- & --- & \textbf{0.68} & \textbf{0.32} & \textbf{38.48} & \textbf{99.64} & --- & --- & $-$0.01 & $-$0.01 & $+$1.12 & $+$0.08 & --- & --- \\
 & P & $-$0.01 & $-$0.07 & $+$0.82 & $+$9.73 & $+$0.00 & $+$0.00 & \textbf{0.99} & \textbf{0.85} & \textbf{1.84} & \textbf{29.21} & \textbf{1.00} & \textbf{1.00} & $+$0.00 & $+$0.03 & $-$0.26 & $-$3.77 & $+$0.00 & $+$0.00 \\
 & UCF & $-$0.01 & $-$0.01 & $+$0.17 & \phantom{+}0.00 & $-$0.02 & $-$0.03 & \textbf{0.20} & \textbf{0.26} & \textbf{99.40} & \textbf{100.00} & \textbf{0.01} & \textbf{0.02} & $+$0.01 & $-$0.01 & $-$0.11 & \phantom{+}0.00 & $-$0.00 & $+$0.01 \\
 & F & $-$0.01 & $-$0.02 & $+$0.33 & \phantom{+}0.00 & $-$0.02 & $-$0.02 & \textbf{0.57} & \textbf{0.37} & \textbf{93.11} & \textbf{100.00} & \textbf{0.12} & \textbf{0.13} & $-$0.00 & $-$0.00 & $+$0.07 & \phantom{+}0.00 & $-$0.01 & $+$0.00 \\
 & FP & $-$0.04 & $-$0.05 & $+$7.27 & \phantom{+}0.00 & $-$0.11 & $-$0.05 & \textbf{0.84} & \textbf{0.73} & \textbf{74.33} & \textbf{99.98} & \textbf{0.62} & \textbf{0.55} & $+$0.03 & $+$0.03 & $-$5.51 & $-$0.20 & $+$0.08 & $+$0.00 \\
 & LR & $-$0.01 & $-$0.05 & $+$0.05 & \phantom{+}0.00 & $-$0.05 & $-$0.05 & \textbf{0.46} & \textbf{0.43} & \textbf{97.57} & \textbf{100.00} & \textbf{0.20} & \textbf{0.20} & $-$0.00 & $+$0.00 & $+$0.24 & \phantom{+}0.00 & $+$0.01 & $+$0.01 \\
\midrule
\multicolumn{20}{l}{\textit{ER topology}} \\
\midrule
 & RC & $-$0.01 & $-$0.01 & $+$1.24 & $-$0.02 & --- & --- & \textbf{0.68} & \textbf{0.32} & \textbf{38.42} & \textbf{99.72} & --- & --- & $-$0.01 & $-$0.01 & $+$1.31 & $-$0.08 & --- & --- \\
 & P & $-$0.01 & $-$0.06 & $+$0.80 & $+$8.77 & $+$0.00 & $+$0.00 & \textbf{0.99} & \textbf{0.85} & \textbf{1.80} & \textbf{28.91} & \textbf{1.00} & \textbf{1.00} & $+$0.00 & $+$0.03 & $-$0.19 & $-$3.25 & $+$0.00 & $+$0.00 \\
 & UCF & $-$0.01 & $-$0.01 & $+$0.17 & \phantom{+}0.00 & $-$0.03 & $-$0.03 & \textbf{0.20} & \textbf{0.26} & \textbf{99.40} & \textbf{100.00} & \textbf{0.01} & \textbf{0.02} & $+$0.01 & $-$0.01 & $-$0.12 & \phantom{+}0.00 & $-$0.00 & $+$0.00 \\
 & F & $-$0.01 & $-$0.01 & $+$0.22 & \phantom{+}0.00 & $-$0.00 & $-$0.00 & \textbf{0.50} & \textbf{0.28} & \textbf{97.38} & \textbf{100.00} & \textbf{0.02} & \textbf{0.02} & $-$0.00 & $-$0.01 & $+$0.22 & \phantom{+}0.00 & $-$0.00 & $+$0.00 \\
 & FP & $-$0.07 & $-$0.03 & $+$6.73 & \phantom{+}0.00 & $-$0.13 & $-$0.07 & \textbf{0.77} & \textbf{0.51} & \textbf{85.01} & \textbf{100.00} & \textbf{0.74} & \textbf{0.68} & $+$0.04 & $+$0.02 & $-$4.61 & \phantom{+}0.00 & $+$0.06 & $-$0.01 \\
 & LR & $+$0.01 & $-$0.00 & $+$0.07 & \phantom{+}0.00 & $-$0.02 & $-$0.02 & \textbf{0.29} & \textbf{0.28} & \textbf{99.25} & \textbf{100.00} & \textbf{0.05} & \textbf{0.05} & $-$0.00 & $+$0.01 & $-$0.01 & \phantom{+}0.00 & $+$0.01 & $+$0.00 \\
\midrule
\multicolumn{20}{l}{\textit{PC topology}} \\
\midrule
 & RC & $+$0.00 & $+$0.00 & $+$0.03 & $-$0.02 & --- & --- & \textbf{0.67} & \textbf{0.31} & \textbf{39.72} & \textbf{99.70} & --- & --- & $+$0.00 & $+$0.01 & $-$0.20 & $+$0.02 & --- & --- \\
 & P & $-$0.01 & $-$0.07 & $+$0.76 & $+$10.20 & $+$0.00 & $+$0.00 & \textbf{0.99} & \textbf{0.85} & \textbf{1.90} & \textbf{28.90} & \textbf{1.00} & \textbf{1.00} & $+$0.00 & $+$0.03 & $-$0.31 & $-$3.20 & $+$0.00 & $+$0.00 \\
 & UCF & $-$0.01 & $-$0.00 & $+$0.20 & \phantom{+}0.00 & $-$0.02 & $-$0.02 & \textbf{0.20} & \textbf{0.25} & \textbf{99.41} & \textbf{100.00} & \textbf{0.01} & \textbf{0.02} & $+$0.01 & $+$0.00 & $-$0.12 & \phantom{+}0.00 & $+$0.00 & $+$0.01 \\
 & F & $-$0.01 & $-$0.01 & $+$0.00 & \phantom{+}0.00 & $-$0.02 & $-$0.02 & \textbf{0.59} & \textbf{0.39} & \textbf{92.55} & \textbf{100.00} & \textbf{0.14} & \textbf{0.14} & $+$0.00 & $+$0.00 & $-$0.31 & \phantom{+}0.00 & $-$0.01 & $+$0.01 \\
 & FP & $-$0.04 & $-$0.04 & $+$7.69 & $+$0.09 & $-$0.11 & $-$0.04 & \textbf{0.86} & \textbf{0.79} & \textbf{70.25} & \textbf{99.91} & \textbf{0.60} & \textbf{0.53} & $+$0.03 & $+$0.03 & $-$7.41 & $-$0.03 & $+$0.08 & $+$0.01 \\
 & LR & $-$0.00 & $-$0.04 & $-$0.30 & \phantom{+}0.00 & $-$0.04 & $-$0.03 & \textbf{0.49} & \textbf{0.46} & \textbf{97.49} & \textbf{100.00} & \textbf{0.20} & \textbf{0.20} & $+$0.01 & $+$0.02 & $-$0.16 & \phantom{+}0.00 & $+$0.02 & $+$0.02 \\
\bottomrule
\end{tabular}}
\end{table*}

\begin{table*}[t]
\centering
\caption{Sensitivity of Gini concentration, recommendation coverage, and popularity-reinforcement AUC to simulation duration $D$. Non-reference columns show $\Delta$ vs.\ the block's reference setting; the reference column shows the absolute value in \textbf{bold}. $G_p$/$G_a$: content-/creator-level Gini, unitless $[0,1]$; $C_p$/$C_a$: content-/creator-level coverage (\%); AUC$_\rho$: interval-normalized area under the popularity$\to$future-exposure Spearman curve.}
\label{tab:sensitivity-summary-duration}
\resizebox{\textwidth}{!}{%
\small
\renewcommand{\arraystretch}{0.9}
\setlength{\tabcolsep}{4pt}
\begin{tabular}{ll r r r r r r r r r r r r r r r}
\toprule
 &  & \multicolumn{5}{c}{$D=30$ \textit{(ref.)}} & \multicolumn{5}{c}{$D=60$ } & \multicolumn{5}{c}{$D=90$ } \\
\cmidrule(lr){3-7}\cmidrule(lr){8-12}\cmidrule(lr){13-17}
Topology & Recommender & $G_p$ & $G_a$ & $C_p$ (\%) & $C_a$ (\%) & AUC$_\rho$ & $\Delta G_p$ & $\Delta G_a$ & $\Delta C_p$ (pp) & $\Delta C_a$ (pp) & $\Delta$AUC$_\rho$ & $\Delta G_p$ & $\Delta G_a$ & $\Delta C_p$ (pp) & $\Delta C_a$ (pp) & $\Delta$AUC$_\rho$ \\
\midrule
\multicolumn{17}{l}{\textit{BA topology}} \\
\midrule
 & RC & \textbf{0.68} & \textbf{0.32} & \textbf{38.48} & \textbf{99.64} & --- & $-$0.01 & $-$0.07 & $+$1.12 & $+$0.36 & --- & $-$0.01 & $-$0.09 & $+$1.16 & $+$0.36 & --- \\
 & P & \textbf{0.99} & \textbf{0.85} & \textbf{1.84} & \textbf{29.21} & \textbf{1.00} & $-$0.00 & $-$0.12 & $+$0.03 & $+$21.09 & $-$0.00 & $-$0.00 & $-$0.21 & $+$0.05 & $+$37.33 & $-$0.00 \\
 & UCF & \textbf{0.20} & \textbf{0.26} & \textbf{99.40} & \textbf{100.00} & \textbf{0.01} & $-$0.01 & $-$0.04 & $+$0.32 & \phantom{+}0.00 & $-$0.01 & $-$0.02 & $-$0.05 & $+$0.40 & \phantom{+}0.00 & $-$0.01 \\
 & F & \textbf{0.57} & \textbf{0.37} & \textbf{93.11} & \textbf{100.00} & \textbf{0.12} & $-$0.01 & $-$0.05 & $+$0.50 & \phantom{+}0.00 & $-$0.00 & $-$0.01 & $-$0.06 & $+$0.56 & \phantom{+}0.00 & $-$0.00 \\
 & FP & \textbf{0.84} & \textbf{0.73} & \textbf{74.33} & \textbf{99.98} & \textbf{0.62} & $-$0.01 & $-$0.02 & $+$0.98 & \phantom{+}0.00 & $-$0.01 & $-$0.01 & $-$0.02 & $+$1.07 & \phantom{+}0.00 & $-$0.00 \\
 & LR & \textbf{0.46} & \textbf{0.43} & \textbf{97.57} & \textbf{100.00} & \textbf{0.20} & $-$0.02 & $-$0.02 & $+$0.55 & \phantom{+}0.00 & $-$0.00 & $+$0.03 & $-$0.04 & $-$7.67 & \phantom{+}0.00 & $-$0.00 \\
\midrule
\multicolumn{17}{l}{\textit{ER topology}} \\
\midrule
 & RC & \textbf{0.68} & \textbf{0.32} & \textbf{38.42} & \textbf{99.72} & --- & $-$0.01 & $-$0.06 & $+$1.14 & $+$0.28 & --- & $-$0.01 & $-$0.08 & $+$1.22 & $+$0.28 & --- \\
 & P & \textbf{0.99} & \textbf{0.85} & \textbf{1.80} & \textbf{28.91} & \textbf{1.00} & $-$0.00 & $-$0.12 & $+$0.08 & $+$21.89 & $-$0.00 & $-$0.00 & $-$0.20 & $+$0.09 & $+$36.89 & $-$0.00 \\
 & UCF & \textbf{0.20} & \textbf{0.26} & \textbf{99.40} & \textbf{100.00} & \textbf{0.01} & $-$0.01 & $-$0.03 & $+$0.29 & \phantom{+}0.00 & $-$0.01 & $-$0.02 & $-$0.04 & $+$0.39 & \phantom{+}0.00 & $-$0.01 \\
 & F & \textbf{0.50} & \textbf{0.28} & \textbf{97.38} & \textbf{100.00} & \textbf{0.02} & $-$0.01 & $-$0.05 & $+$0.48 & \phantom{+}0.00 & $+$0.00 & $-$0.01 & $-$0.07 & $+$0.50 & \phantom{+}0.00 & $-$0.00 \\
 & FP & \textbf{0.77} & \textbf{0.51} & \textbf{85.01} & \textbf{100.00} & \textbf{0.74} & $-$0.01 & $-$0.07 & $+$1.26 & \phantom{+}0.00 & $-$0.01 & $-$0.01 & $-$0.11 & $+$1.40 & \phantom{+}0.00 & $-$0.01 \\
 & LR & \textbf{0.29} & \textbf{0.28} & \textbf{99.25} & \textbf{100.00} & \textbf{0.05} & $-$0.01 & $-$0.01 & $+$0.27 & \phantom{+}0.00 & $-$0.00 & $-$0.01 & $-$0.02 & $+$0.32 & \phantom{+}0.00 & $-$0.00 \\
\midrule
\multicolumn{17}{l}{\textit{PC topology}} \\
\midrule
 & RC & \textbf{0.67} & \textbf{0.31} & \textbf{39.72} & \textbf{99.70} & --- & $+$0.00 & $-$0.05 & $-$0.03 & $+$0.30 & --- & $+$0.00 & $-$0.08 & $-$0.05 & $+$0.30 & --- \\
 & P & \textbf{0.99} & \textbf{0.85} & \textbf{1.90} & \textbf{28.90} & \textbf{1.00} & $+$0.00 & $-$0.13 & $-$0.00 & $+$22.56 & $-$0.00 & $+$0.00 & $-$0.20 & $-$0.04 & $+$36.28 & $-$0.00 \\
 & UCF & \textbf{0.20} & \textbf{0.25} & \textbf{99.41} & \textbf{100.00} & \textbf{0.01} & $-$0.01 & $-$0.03 & $+$0.29 & \phantom{+}0.00 & $-$0.00 & $-$0.02 & $-$0.04 & $+$0.39 & \phantom{+}0.00 & $-$0.00 \\
 & F & \textbf{0.59} & \textbf{0.39} & \textbf{92.55} & \textbf{100.00} & \textbf{0.14} & $-$0.00 & $-$0.03 & $+$0.23 & \phantom{+}0.00 & $-$0.00 & $+$0.00 & $-$0.05 & $+$0.34 & \phantom{+}0.00 & $+$0.00 \\
 & FP & \textbf{0.86} & \textbf{0.79} & \textbf{70.25} & \textbf{99.91} & \textbf{0.60} & $-$0.00 & $-$0.01 & $+$0.12 & $+$0.09 & $-$0.00 & $-$0.00 & $-$0.01 & $+$0.14 & $+$0.09 & $-$0.00 \\
 & LR & \textbf{0.49} & \textbf{0.46} & \textbf{97.49} & \textbf{100.00} & \textbf{0.20} & $-$0.00 & $-$0.01 & $+$0.21 & \phantom{+}0.00 & $+$0.00 & $-$0.00 & $-$0.01 & $+$0.28 & \phantom{+}0.00 & $+$0.01 \\
\bottomrule
\end{tabular}}
\end{table*}

\begin{table*}[h!]
\centering
\caption{Sensitivity of Gini concentration, recommendation coverage, and popularity-reinforcement AUC to feed size $k$. Non-reference columns show $\Delta$ vs.\ the block's reference setting; the reference column shows the absolute value in \textbf{bold}. $G_p$/$G_a$: content-/creator-level Gini, unitless $[0,1]$; $C_p$/$C_a$: content-/creator-level coverage (\%); AUC$_\rho$: interval-normalized area under the popularity$\to$future-exposure Spearman curve.}
\label{tab:sensitivity-summary-feed}
\resizebox{\textwidth}{!}{%
\small
\renewcommand{\arraystretch}{0.9}
\setlength{\tabcolsep}{4pt}
\begin{tabular}{ll r r r r r r r r r r r r r r r}
\toprule
 &  & \multicolumn{5}{c}{$k=6$ } & \multicolumn{5}{c}{$k=12$ \textit{(ref.)}} & \multicolumn{5}{c}{$k=24$ } \\
\cmidrule(lr){3-7}\cmidrule(lr){8-12}\cmidrule(lr){13-17}
Topology & Recommender & $\Delta G_p$ & $\Delta G_a$ & $\Delta C_p$ (pp) & $\Delta C_a$ (pp) & $\Delta$AUC$_\rho$ & $G_p$ & $G_a$ & $C_p$ (\%) & $C_a$ (\%) & AUC$_\rho$ & $\Delta G_p$ & $\Delta G_a$ & $\Delta C_p$ (pp) & $\Delta C_a$ (pp) & $\Delta$AUC$_\rho$ \\
\midrule
\multicolumn{17}{l}{\textit{BA topology}} \\
\midrule
 & RC & $+$0.16 & $+$0.07 & $-$18.62 & $-$3.82 & --- & \textbf{0.68} & \textbf{0.32} & \textbf{38.48} & \textbf{99.64} & --- & $-$0.32 & $-$0.06 & $+$37.90 & $+$0.36 & --- \\
 & P & $+$0.01 & $+$0.06 & $-$1.00 & $-$14.11 & $+$0.00 & \textbf{0.99} & \textbf{0.85} & \textbf{1.84} & \textbf{29.21} & \textbf{1.00} & $-$0.01 & $-$0.11 & $+$2.90 & $+$26.57 & $-$0.00 \\
 & UCF & $+$0.07 & $-$0.01 & $-$1.02 & \phantom{+}0.00 & $+$0.00 & \textbf{0.20} & \textbf{0.26} & \textbf{99.40} & \textbf{100.00} & \textbf{0.01} & $-$0.05 & $-$0.01 & $+$0.26 & \phantom{+}0.00 & $+$0.00 \\
 & F & $+$0.13 & $+$0.01 & $-$19.58 & \phantom{+}0.00 & $-$0.03 & \textbf{0.57} & \textbf{0.37} & \textbf{93.11} & \textbf{100.00} & \textbf{0.12} & $-$0.15 & $-$0.04 & $+$5.97 & \phantom{+}0.00 & $-$0.02 \\
 & FP & $+$0.10 & $+$0.08 & $-$36.73 & $-$1.62 & $+$0.12 & \textbf{0.84} & \textbf{0.73} & \textbf{74.33} & \textbf{99.98} & \textbf{0.62} & $-$0.08 & $-$0.08 & $+$21.00 & $+$0.02 & $-$0.15 \\
 & LR & $+$0.06 & $+$0.02 & $-$8.06 & \phantom{+}0.00 & $-$0.01 & \textbf{0.46} & \textbf{0.43} & \textbf{97.57} & \textbf{100.00} & \textbf{0.20} & $-$0.07 & $-$0.07 & $+$0.89 & \phantom{+}0.00 & $-$0.05 \\
\midrule
\multicolumn{17}{l}{\textit{ER topology}} \\
\midrule
 & RC & $+$0.15 & $+$0.07 & $-$18.55 & $-$4.34 & --- & \textbf{0.68} & \textbf{0.32} & \textbf{38.42} & \textbf{99.72} & --- & $-$0.32 & $-$0.06 & $+$37.98 & $+$0.28 & --- \\
 & P & $+$0.01 & $+$0.06 & $-$0.94 & $-$13.43 & $+$0.00 & \textbf{0.99} & \textbf{0.85} & \textbf{1.80} & \textbf{28.91} & \textbf{1.00} & $-$0.01 & $-$0.11 & $+$2.99 & $+$26.51 & $-$0.00 \\
 & UCF & $+$0.07 & $-$0.00 & $-$1.00 & \phantom{+}0.00 & $-$0.00 & \textbf{0.20} & \textbf{0.26} & \textbf{99.40} & \textbf{100.00} & \textbf{0.01} & $-$0.04 & $-$0.01 & $+$0.24 & \phantom{+}0.00 & $+$0.00 \\
 & F & $+$0.15 & $+$0.03 & $-$15.27 & \phantom{+}0.00 & $+$0.00 & \textbf{0.50} & \textbf{0.28} & \textbf{97.38} & \textbf{100.00} & \textbf{0.02} & $-$0.13 & $-$0.02 & $+$2.21 & \phantom{+}0.00 & $+$0.00 \\
 & FP & $+$0.14 & $+$0.10 & $-$39.99 & $-$0.04 & $+$0.10 & \textbf{0.77} & \textbf{0.51} & \textbf{85.01} & \textbf{100.00} & \textbf{0.74} & $-$0.14 & $-$0.05 & $+$14.01 & \phantom{+}0.00 & $-$0.17 \\
 & LR & $+$0.08 & $+$0.01 & $-$4.08 & \phantom{+}0.00 & $+$0.01 & \textbf{0.29} & \textbf{0.28} & \textbf{99.25} & \textbf{100.00} & \textbf{0.05} & $-$0.03 & $-$0.01 & $+$0.40 & \phantom{+}0.00 & $-$0.01 \\
\midrule
\multicolumn{17}{l}{\textit{PC topology}} \\
\midrule
 & RC & $+$0.17 & $+$0.07 & $-$19.87 & $-$4.40 & --- & \textbf{0.67} & \textbf{0.31} & \textbf{39.72} & \textbf{99.70} & --- & $-$0.31 & $-$0.04 & $+$36.92 & $+$0.30 & --- \\
 & P & $+$0.01 & $+$0.06 & $-$1.03 & $-$13.12 & $+$0.00 & \textbf{0.99} & \textbf{0.85} & \textbf{1.90} & \textbf{28.90} & \textbf{1.00} & $-$0.01 & $-$0.10 & $+$2.79 & $+$26.12 & $-$0.00 \\
 & UCF & $+$0.07 & $-$0.00 & $-$0.97 & \phantom{+}0.00 & $-$0.00 & \textbf{0.20} & \textbf{0.25} & \textbf{99.41} & \textbf{100.00} & \textbf{0.01} & $-$0.05 & $-$0.00 & $+$0.27 & \phantom{+}0.00 & $+$0.00 \\
 & F & $+$0.13 & $+$0.02 & $-$21.74 & \phantom{+}0.00 & $-$0.02 & \textbf{0.59} & \textbf{0.39} & \textbf{92.55} & \textbf{100.00} & \textbf{0.14} & $-$0.14 & $-$0.04 & $+$6.34 & \phantom{+}0.00 & $-$0.01 \\
 & FP & $+$0.09 & $+$0.08 & $-$37.45 & $-$2.51 & $+$0.12 & \textbf{0.86} & \textbf{0.79} & \textbf{70.25} & \textbf{99.91} & \textbf{0.60} & $-$0.08 & $-$0.08 & $+$24.11 & $+$0.09 & $-$0.14 \\
 & LR & $+$0.07 & $+$0.04 & $-$9.92 & $-$0.02 & $+$0.01 & \textbf{0.49} & \textbf{0.46} & \textbf{97.49} & \textbf{100.00} & \textbf{0.20} & $-$0.08 & $-$0.08 & $+$1.95 & \phantom{+}0.00 & $-$0.06 \\
\bottomrule
\end{tabular}}
\end{table*}

\begin{table*}[h!]
\centering
\caption{Sensitivity of Gini concentration, recommendation coverage, and popularity-reinforcement AUC to population size $N$. Non-reference columns show $\Delta$ vs.\ the block's reference setting; the reference column shows the absolute value in \textbf{bold}. $G_p$/$G_a$: content-/creator-level Gini, unitless $[0,1]$; $C_p$/$C_a$: content-/creator-level coverage (\%); AUC$_\rho$: interval-normalized area under the popularity$\to$future-exposure Spearman curve.}
\label{tab:sensitivity-summary-population}
\resizebox{\textwidth}{!}{%
\small
\renewcommand{\arraystretch}{0.9}
\setlength{\tabcolsep}{4pt}
\begin{tabular}{ll r r r r r r r r r r r r r r r}
\toprule
 &  & \multicolumn{5}{c}{$N=1000$ \textit{(ref.)}} & \multicolumn{5}{c}{$N=5000$ } & \multicolumn{5}{c}{$N=10000$ } \\
\cmidrule(lr){3-7}\cmidrule(lr){8-12}\cmidrule(lr){13-17}
Topology & Recommender & $G_p$ & $G_a$ & $C_p$ (\%) & $C_a$ (\%) & AUC$_\rho$ & $\Delta G_p$ & $\Delta G_a$ & $\Delta C_p$ (pp) & $\Delta C_a$ (pp) & $\Delta$AUC$_\rho$ & $\Delta G_p$ & $\Delta G_a$ & $\Delta C_p$ (pp) & $\Delta C_a$ (pp) & $\Delta$AUC$_\rho$ \\
\midrule
\multicolumn{17}{l}{\textit{BA topology}} \\
\midrule
 & RC & \textbf{0.68} & \textbf{0.32} & \textbf{38.48} & \textbf{99.64} & --- & $+$0.25 & $+$0.25 & $-$30.72 & $-$30.16 & --- & $+$0.29 & $+$0.39 & $-$34.59 & $-$55.05 & --- \\
 & P & \textbf{0.99} & \textbf{0.85} & \textbf{1.84} & \textbf{29.21} & \textbf{1.00} & $+$0.01 & $+$0.12 & $-$1.51 & $-$23.51 & $+$0.00 & $+$0.01 & $+$0.14 & $-$1.68 & $-$26.40 & $-$0.00 \\
 & UCF & \textbf{0.20} & \textbf{0.26} & \textbf{99.40} & \textbf{100.00} & \textbf{0.01} & $+$0.00 & $-$0.00 & $+$0.02 & \phantom{+}0.00 & $+$0.00 & $+$0.00 & $-$0.01 & $-$0.00 & \phantom{+}0.00 & $-$0.00 \\
 & F & \textbf{0.57} & \textbf{0.37} & \textbf{93.11} & \textbf{100.00} & \textbf{0.12} & $+$0.11 & $+$0.11 & $-$0.80 & \phantom{+}0.00 & $-$0.05 & $+$0.16 & $+$0.21 & $-$4.21 & $-$0.00 & $-$0.05 \\
 & FP & \textbf{0.84} & \textbf{0.73} & \textbf{74.33} & \textbf{99.98} & \textbf{0.62} & $+$0.13 & $+$0.14 & $-$37.52 & $-$1.42 & $+$0.32 & $+$0.15 & $+$0.20 & $-$53.31 & $-$11.54 & $+$0.36 \\
 & LR & \textbf{0.46} & \textbf{0.43} & \textbf{97.57} & \textbf{100.00} & \textbf{0.20} & $-$0.03 & $-$0.01 & $+$1.22 & \phantom{+}0.00 & $-$0.06 & $-$0.03 & $-$0.02 & $+$1.12 & \phantom{+}0.00 & $-$0.10 \\
\midrule
\multicolumn{17}{l}{\textit{ER topology}} \\
\midrule
 & RC & \textbf{0.68} & \textbf{0.32} & \textbf{38.42} & \textbf{99.72} & --- & $+$0.25 & $+$0.25 & $-$30.64 & $-$30.15 & --- & $+$0.29 & $+$0.39 & $-$34.54 & $-$54.95 & --- \\
 & P & \textbf{0.99} & \textbf{0.85} & \textbf{1.80} & \textbf{28.91} & \textbf{1.00} & $+$0.01 & $+$0.12 & $-$1.47 & $-$23.07 & $+$0.00 & $+$0.01 & $+$0.14 & $-$1.63 & $-$25.84 & $+$0.00 \\
 & UCF & \textbf{0.20} & \textbf{0.26} & \textbf{99.40} & \textbf{100.00} & \textbf{0.01} & $+$0.00 & $-$0.00 & $+$0.01 & \phantom{+}0.00 & $+$0.00 & $+$0.00 & $-$0.00 & $+$0.00 & \phantom{+}0.00 & $-$0.00 \\
 & F & \textbf{0.50} & \textbf{0.28} & \textbf{97.38} & \textbf{100.00} & \textbf{0.02} & $+$0.08 & $+$0.10 & $+$0.66 & \phantom{+}0.00 & $-$0.02 & $+$0.10 & $+$0.18 & $-$0.58 & $-$0.00 & $-$0.02 \\
 & FP & \textbf{0.77} & \textbf{0.51} & \textbf{85.01} & \textbf{100.00} & \textbf{0.74} & $+$0.17 & $+$0.14 & $-$38.69 & $-$0.12 & $+$0.23 & $+$0.20 & $+$0.25 & $-$55.88 & $-$4.35 & $+$0.25 \\
 & LR & \textbf{0.29} & \textbf{0.28} & \textbf{99.25} & \textbf{100.00} & \textbf{0.05} & $-$0.04 & $-$0.01 & $+$0.46 & \phantom{+}0.00 & $-$0.02 & $-$0.03 & $-$0.02 & $+$0.44 & \phantom{+}0.00 & $-$0.03 \\
\midrule
\multicolumn{17}{l}{\textit{PC topology}} \\
\midrule
 & RC & \textbf{0.67} & \textbf{0.31} & \textbf{39.72} & \textbf{99.70} & --- & $+$0.27 & $+$0.26 & $-$31.94 & $-$30.59 & --- & $+$0.30 & $+$0.41 & $-$35.84 & $-$55.20 & --- \\
 & P & \textbf{0.99} & \textbf{0.85} & \textbf{1.90} & \textbf{28.90} & \textbf{1.00} & $+$0.01 & $+$0.12 & $-$1.56 & $-$22.90 & $-$0.00 & $+$0.01 & $+$0.14 & $-$1.73 & $-$25.89 & $-$0.00 \\
 & UCF & \textbf{0.20} & \textbf{0.25} & \textbf{99.41} & \textbf{100.00} & \textbf{0.01} & $+$0.00 & $+$0.01 & $+$0.00 & \phantom{+}0.00 & $+$0.00 & $-$0.00 & $+$0.00 & $+$0.01 & \phantom{+}0.00 & $+$0.00 \\
 & F & \textbf{0.59} & \textbf{0.39} & \textbf{92.55} & \textbf{100.00} & \textbf{0.14} & $+$0.12 & $+$0.13 & $-$2.03 & \phantom{+}0.00 & $-$0.04 & $+$0.18 & $+$0.23 & $-$6.30 & \phantom{+}0.00 & $-$0.06 \\
 & FP & \textbf{0.86} & \textbf{0.79} & \textbf{70.25} & \textbf{99.91} & \textbf{0.60} & $+$0.12 & $+$0.14 & $-$42.69 & $-$4.24 & $+$0.33 & $+$0.13 & $+$0.19 & $-$57.91 & $-$27.02 & $+$0.37 \\
 & LR & \textbf{0.49} & \textbf{0.46} & \textbf{97.49} & \textbf{100.00} & \textbf{0.20} & $+$0.00 & $+$0.02 & $+$0.71 & \phantom{+}0.00 & $-$0.02 & $+$0.03 & $+$0.04 & $+$0.29 & \phantom{+}0.00 & $-$0.05 \\
\bottomrule
\end{tabular}}
\end{table*}

\end{document}